\documentclass[%
superscriptaddress,
 amsmath,amssymb,
prb,
floatfix,
]{revtex4-2}

\usepackage[utf8]{inputenc}
\usepackage[T1]{fontenc}
\usepackage{graphicx}% Include figure files
\usepackage{dcolumn}% Align table columns on decimal point
\usepackage{bm}% bold math
\usepackage{hyperref}% add hypertext capabilities
\hypersetup{colorlinks=true,linkcolor=blue,citecolor=blue,urlcolor=blue}
\begin{document}

\preprint{}

\title{Unified Statistical Theory of Heat Conduction in Nonuniform Media} %{Beyond Fourier’s Law and the Kapitza Model of Thermal Interfaces}

\author{Yi Zeng}
 \email{Contact author: yi.zeng@nlr.gov}
 %\altaffiliation[Also at ]{National Renewable Energy Laboratory, Golden, CO, USA.}%Lines break automatically or can be forced with \\
 \affiliation{%
 National Laboratory of the Rockies, Golden, CO 80401, USA.
}%

\author{Jianjun Dong}%
 \email{Contact author: dongjia@auburn.edu}
\affiliation{%
Department of Physics, Auburn University, Auburn, AL 36849, USA.
}%

%  \homepage{http://www.Second.institution.edu/~Charlie.Author}
% \affiliation{
%  First affiliation for this author
% }%
% \affiliation{
%  second institution for this author
% \affiliation{%
%  Authors' institution and/or address\\
%  This line break forced with \textbackslash\textbackslash
% }%

% \collaboration{CLEO Collaboration}%\noaffiliation

\date{\today}% It is always \today, today,

\begin{abstract}
Using the Zwanzig projection-operator formalism, we derive a causal two-point spatiotemporal kernel for heat conduction, defined microscopically as a space-resolved equilibrium heat-flux time-correlation function, that encodes temporal memory, spatial nonlocality, and material heterogeneity on equal footing. Classical diffusion, nonlocal transport, and hydrodynamic models emerge as controlled asymptotic limits of this kernel, providing a unified constitutive description across diffusive, quasi-ballistic, and hydrodynamic regimes. Interfacial heat transfer is incorporated through a spatially resolved kernel formulation, in which the conventional Kapitza resistance arises as a coarse-grained limit. The kernel admits a spatiotemporal Green--Kubo representation and can, in principle, be evaluated from atomistic simulations for bulk media, providing a direct connection between microscopic dynamics and continuum transport without empirical closure. For crystalline solids, we derive explicit kernel forms in the hydrodynamic and attenuated-streaming limits and introduce a hybrid reduction that captures the coexistence of collective and quasi-ballistic transport. For disordered harmonic solids, the framework recovers a spatial diffusion kernel consistent with the Allen--Feldman limit. To illustrate the theory, we construct the kernel for silicon at room temperature within the relaxation-time approximation and apply it to transient thermal grating configurations. Spatial nonlocality associated with the phonon mean-free-path distribution is the primary source of deviation from Fourier transport under these conditions, while temporal memory mainly influences short-time dynamics. These findings identify the spatiotemporal kernel as a unifying constitutive descriptor whose coarse-grained limits recover conventional transport coefficients.
\end{abstract}

% For cover letter: We introduce a unified, first-principles framework for heat conduction in which bulk, nanoscale, and interfacial transport are described by a single spatiotemporal kernel, with all conventional models—including Kapitza resistance—emerging as controlled asymptotic limits.

% \begin{description}
% Secondary publications and information retrieval purposes.
% You may use the \texttt{description} environment to structure your abstract;
% use the optional argument of the \verb+\item+ command to give the category of each item. 
% \end{description}

%Use showkeys class option if keyword
\maketitle

%\tableofcontents
\newpage
\section{Introduction}
\label{sec:introduction}

Heat conduction is conventionally described by combining local energy conservation with Fourier's law, which assumes that the heat flux responds locally and instantaneously to the temperature gradient. This description, which implicitly relies on a coarse-grained limit in which microscopic correlations decay sufficiently rapidly, has been highly successful in macroscopic regimes, where transport can be characterized by bulk coefficients such as the thermal conductivity. Within statistical mechanics, this picture is supported by the Green--Kubo relation, which expresses the bulk conductivity as the time integral of equilibrium heat-flux autocorrelation functions~\cite{dong2001_PhysRevLett.86.2361}.

Despite its success, the Fourier framework rests on two key assumptions: instantaneous temporal response and spatial locality. These assumptions may no longer hold at ultrafast time scales and nanometer length scales, where intrinsic relaxation times and phonon mean free paths (MFP) become comparable to experimental time scales and device dimensions. A growing body of experiments has reported clear deviations from Fourier behavior, including quasi-ballistic transport~\cite{maznev2011onset,minnich2011thermal,johnson2013direct,johnson2015non,hua2019generalized,beardo2021general}, wave-like temperature propagation (second sound)~\cite{ackerman1966second,mcnelly1970heat,jackson1970second,pohl1976observation,narayanamurti1972observation,danil1979observation,huberman2019observation,ding2022observation,jeong2021transient,beardo2021observation}, and signatures of Poiseuille flow~\cite{machida2018observation,markov2018hydrodynamic,martelli2018thermal,machida2020phonon,huang2023observation,huang2024graphite}.
These phenomena are not only of fundamental interest but are increasingly relevant for applications. In modern integrated circuits, device dimensions approach the sub-10~nm scale and operating frequencies reach the sub-terahertz regime, overlapping with intrinsic phonon transport scales in typical semiconductors.

To account for these non-Fourier effects, a range of generalized transport models have been developed. The Maxwell--Cattaneo--Vernotte (MCV) equation introduces a finite relaxation time to capture temporal memory~\cite{maxwell1866dynamical,cattaneo1948atti,vernotte1958paradoxes,chester1963second}, while the Guyer--Krumhansl (G--K) equation incorporates higher-order spatial gradients and provides a unified description of nonlocal and hydrodynamic transport~\cite{guyer1966solution,guyer1966thermal,joseph1989heat,simoncelli2020generalization,sendra2022hydrodynamic,guo2018phonon}. The G--K equation can be rigorously derived from the phonon Boltzmann transport equation (phBTE) and has been successfully applied to interpret experimental observations across a range of materials and geometries~\cite{sendra2021derivation,beardo2022hydrodynamic,xiang2022time}. However, these reduced models are intrinsically tied to specific asymptotic regimes whose validity relies on a separation of characteristic time or length scales. In situations where multiple transport mechanisms coexist, such as in materials with broad mean-free-path distributions, strong disorder, or complex interfaces, no single reduced description provides a systematically controlled representation across regimes.

At the microscopic level, first-principles solutions of the phBTE provide a rigorous description in which transport emerges from phonon dispersions and scattering processes~\cite{hua2014analytical,hua_minnich_2018_PRB,chiloyan2021green,malviya2025efficient}. This formulation is intrinsically tied to the phonon quasiparticle picture and is therefore most naturally applicable to crystalline solids with well-defined phonon modes and group velocities. Its extension to non-crystalline materials is not direct and requires alternative definitions of energy-carrying excitations. Moreover, even in crystalline systems, practical implementations are best suited to homogeneous or weakly perturbed conditions; extension to spatially inhomogeneous or interfacial problems is computationally demanding and often leads to geometry-specific descriptions that lack compact, transferable constitutive form.

In contrast, the Green--Kubo formalism provides a general statistical framework applicable to all materials, expressing thermal transport in terms of equilibrium heat-flux correlations. However, in its conventional form it yields only bulk transport coefficients, corresponding to a fully coarse-grained limit. Recent work has extended this approach to transient transport through a time-domain kernel under the assumption of spatial locality~\cite{zeng_and_dong_2019_vibFPE,crawford2024theory}, and to interfacial heat exchange via a non-Markovian memory kernel~\cite{Zeng2021ITC}, enabling descriptions of temporal memory while leaving spatial nonlocality unresolved. Taken together, these approaches reveal a fundamental gap: microscopic theories provide rigor but limited transferability, while reduced models offer compact descriptions whose domains of validity must be established separately. More fundamentally, these limitations reflect a deeper issue: the constitutive description of heat transport is typically expressed in terms of local or low-order parameters, whereas the underlying microscopic response is inherently nonlocal in both space and time.  A unified framework that retains both temporal and spatial structure, while systematically connecting microscopic dynamics to macroscopic transport across regimes, remains lacking.

In this work, we develop such a framework based on the Zwanzig projection-operator formalism. Within this approach, heat conduction is described by a causal two-point spatiotemporal kernel $\overleftrightarrow{\mathbb{Z}}(\vec r,\vec R,t)$ that relates the heat flux to the history of the temperature gradient over space and time. This kernel encodes temporal memory, spatial nonlocality, and material heterogeneity on equal footing. Classical and non-Fourier transport models emerge as controlled asymptotic reductions of this single constitutive object, eliminating the need for regime-specific model selection. In particular, the G--K equation arises as a hydrodynamic limit associated with mode separation, while quasi-ballistic transport is captured through nonlocal streaming contributions. Interfacial heat transfer is described within the same framework, with the Kapitza resistance emerging as a coarse-grained limit of a spatially resolved kernel. The kernel admits a microscopic definition through a spatiotemporal Green--Kubo (stGK) relation, expressed as a space-resolved equilibrium heat-flux time-correlation function. Because it is defined in terms of equilibrium fluctuations, it can be evaluated from atomistic simulations without empirical closure parameters and applied to crystalline, disordered, and heterogeneous systems. For disordered harmonic solids, the formalism recovers a spatial diffusion kernel consistent with the Allen--Feldman limit~\cite{Allen_Feldman_PhysRevB.48.12581}, demonstrating applicability beyond the phonon-gas picture.

The key result of this work is the identification of the heat conduction kernel as a space-projected heat-flux correlation function, which generalizes the Green--Kubo relation to nonuniform systems. This establishes a direct connection between microscopic fluctuations and nonlocal constitutive relations without invoking phenomenological boundary conditions. Specifically, the present work (i) identifies the two-point kernel $\overleftrightarrow{\mathbb{Z}}(\vec{r},\vec{R},t)$ as the primary constitutive object, placing spatial nonlocality and temporal memory on equal footing; (ii) derives a generalized Green--Kubo relation that resolves heat-flux correlations in both space and time; (iii) demonstrates that classical diffusion, nonlocal transport, and hydrodynamic regimes emerge as controlled asymptotic limits of the same kernel; and (iv) formulates interfacial heat transport within this framework, in which Kapitza resistance arises as a reduced descriptor of an underlying spatially resolved kernel. To demonstrate the practical utility of the framework, we evaluate the kernel for bulk silicon at room temperature using first-principles phonon properties within the relaxation-time approximation (RTA) and apply it to simulate transient thermal grating (TTG) experiments across diffusive and quasi-ballistic regimes. By systematically comparing the full kernel with its reduced limits, we show that spatial nonlocality associated with the phonon mean-free-path distribution is the primary source of deviation from Fourier transport, while temporal memory primarily affects short-time dynamics.

The remainder of this paper is organized as follows. Section~II develops the Zwanzig projection-operator based formulation and derives reduced constitutive models as asymptotic limits of the kernel. Section~III establishes the microscopic foundation through the spatiotemporal Green--Kubo relation. Section~IV presents numerical results for bulk silicon, and Section~V summarizes the main conclusions and outlook.

\section{Phonon-Mediated Heat Conduction}
\label{sec:phonon_conduction}

\subsection{Zwanzig-Projected phBTE}
\label{sec:zwanzig}

Using the Zwanzig projection-operator formalism~\cite{zwanzig1960ensemble,zwanzig1961memory,zwanzig1964elementary}, we decompose the microscopic phonon dynamics into three classes of projected variables: a conserved temperature field, current-carrying flux modes, and nonlocal distortion modes (Table~\ref{tab:zwanzig_variables}). Their coupled dynamics make explicit the microscopic origin of temporal memory and spatial nonlocality, and provide the building blocks from which the spatiotemporal kernel is constructed in Sec.~\ref{sec:unified_framework}. Supporting technical details, including causal integral solutions, their inhomogeneous generalization, and the derivations of reduced kernel forms, are collected in Appendix~\ref{app:model-coupled-dynamics}.

\subsubsection{Phonon Boltzmann Transport Equation}
\label{sec:conventional_phBTE}

Within the phBTE, thermal excitations are described by the occupation numbers of phonon normal modes labeled by $\alpha=(\vec q,b)$, where $\vec q$ is the wave vector and $b$ is the branch index. The local occupations $\{n_\alpha(\vec r,t)\}$ specify the space- and time-dependent microscopic state of the phonon system. Nonequilibrium dynamics is described by deviations from global equilibrium, $
\delta n_\alpha(\vec r,t)=n_\alpha(\vec r,t)-n_\alpha^{\mathrm{eq}}(T_0)$,
where $n_\alpha^{\mathrm{eq}}(T_0)$ is the Bose--Einstein distribution at the reference temperature $T_0$. A central approximation is the linearized phonon scattering operator $\mathcal{L}$, which governs intermode scattering near equilibrium. The relaxation dynamics may be written as
\begin{equation}
\label{eq:lpsm}
\left. \frac{d n_{\alpha}(\vec{r},t)}{d t} \right|_{\mathrm{relax}}
\approx
- \sum_{\beta} L_{\alpha \beta}
\frac{\sinh \!\left(\frac{\hbar \omega_{\beta}}{2 k_\mathrm{B} T_0}\right)}
{\sinh \!\left(\frac{\hbar \omega_{\alpha}}{2 k_\mathrm{B} T_0}\right)}
\left[n_{\beta}(\vec{r},t)-n^{\mathrm{eq}}_{\beta}(T_0)\right],
\end{equation}
where $L_{\alpha\beta}$ are the matrix elements of $\mathcal{L}$ in the phonon-mode basis. The operator $\mathcal{L}$ is symmetric and positive semidefinite, ensuring relaxation toward equilibrium.

To incorporate spatial inhomogeneity, we allow phonon properties, including $\mathcal{L}(\vec r)$, $\omega_\alpha(\vec r)$, and $\vec v_\alpha(\vec r)$, to depend weakly on position. This is justified under a \emph{quasi-homogeneous} assumption: material properties vary slowly enough that local phonon modes remain well defined. The lattice may then be treated as locally homogeneous, with phonons propagating through a medium whose properties evolve adiabatically in space. Under this assumption, the linearized, space-resolved phBTE becomes
\begin{equation}
\label{eq:bte_phonon_mode}
\frac{\partial n_{\alpha}(\vec{r},t)}{\partial t}
+ \vec{v}_{\alpha}(\vec{r}) \cdot \nabla n_{\alpha}(\vec{r}, t)
=
- \sum_{\beta} L_{\alpha \beta}(\vec{r})
\frac{\sinh \!\left(\hbar \omega_{\beta}(\vec{r}) / 2 k_\mathrm{B} T_0\right)}
{\sinh \!\left(\hbar \omega_{\alpha}(\vec{r}) / 2 k_\mathrm{B} T_0\right)}
\left[n_{\beta}(\vec{r},t)-n^{\mathrm{eq}}_{\beta}(T_0)\right].
\end{equation}

This quasi-homogeneous formulation provides the microscopic starting point for the projected transport theory developed below. Additional thermodynamic background, including the local thermal-equilibrium definition of temperature and its relation to mode-resolved heat capacity and equilibrium phonon fluctuations, is summarized in Appendix~\ref{app:phonon}.

\subsubsection{Relevant and Irrelevant Variables}
\label{sec:zwanzig_variables}
%%%%%%%%%%%%%%%%%%%%%%%%%%%%%%%%%%%%%%%%%%%%%%%%%%%%%%%%%%
\begin{table}[tb]
\caption{Zwanzig-projected variables and their physical roles in phonon-mediated heat conduction.}
\label{tab:zwanzig_variables}
\centering
\begin{tabular}{c c p{10cm}}
\hline\hline
Symbol & Role & Physical meaning \\
\hline
$T(\vec r,t)$ & Conserved & Local temperature field; energy density \\
$\eta_\lambda(\vec r,t)$ & Odd-parity (flux) & Current-carrying modes reconstructing heat flux $\vec j$ \\
$\theta_\iota(\vec r,t)$ & Even-parity (nonlocal) & $\vec q$-symmetric distortion modes mediating spatial nonlocality \\[3pt]
$\vec{\Lambda}_\lambda$ & Coupling vector & Maps flux mode $\eta_\lambda$ to macroscopic heat flux \\
$\vec{\Pi}_{\lambda\iota}$ & Coupling vector & Couples flux and nonlocal modes; governs nonlocality \\[3pt]
$\gamma_\lambda^{-1}$ & Time scale & Flux-mode relaxation time (temporal memory) \\
$\varepsilon_\iota^{-1}$ & Time scale & Nonlocal-mode relaxation time \\
\hline\hline
\end{tabular}
\end{table}
%%%%%%%%%%%%%%%%%%%%%%%%%%%%%%%%%%%%%%%%%%%%%%%%%%%%%%%%%%

We now reformulate the phBTE using the Zwanzig projection-operator formalism; the projected variables and associated quantities introduced below are summarized in Table~\ref{tab:zwanzig_variables}. The central idea is to partition the microscopic degrees of freedom into a reduced set of slowly evolving \emph{relevant} variables and a complementary set of rapidly relaxing \emph{irrelevant} variables. We take the relevant variable to be the coarse-grained local temperature field $T(\vec r,t)$, which is directly accessible experimentally and encodes the conserved energy density. All remaining phonon degrees of freedom are treated as irrelevant variables whose influence enters through the constitutive dynamics of the heat flux. The construction proceeds by expanding the phonon distribution in the orthonormal eigenbasis of the scattering operator $\mathcal{L}$. The eigenmodes of $\mathcal{L}$ are defined by
\begin{subequations}
\label{eq:eigenmodes}
\begin{align}
&\sum_{\beta} L_{\alpha \beta} \left(\frac{c_\beta}{C}\right)^{1/2} = 0, \\
&\sum_{\beta} L_{\alpha \beta} \nu_{\iota\beta} = \varepsilon_{\iota} \nu_{\iota\alpha}, \\
&\sum_{\beta} L_{\alpha \beta} \chi_{\lambda\beta} = \gamma_{\lambda} \chi_{\lambda\alpha},
\end{align}
\end{subequations}
where $C$ and $c_\alpha$ denote the total and mode-resolved heat capacities, respectively (Appendix~\ref{app:phonon}). This spectrum consists of a zero-eigenvalue mode associated with energy conservation, together with a hierarchy of relaxing modes. The parity classification follows from inversion symmetry of the scattering operator, $L_{\alpha\beta} = L_{\bar\alpha\bar\beta}$ with $\bar\alpha = (-\vec q,b)$, which ensures that the eigenvectors can be chosen with definite parity under $\vec q \leftrightarrow -\vec q$. We therefore separate the relaxing modes into even-parity modes $\{\nu_\iota\}$ and odd-parity modes $\{\chi_\lambda\}$.

The zero-eigenvalue mode corresponds to the conserved temperature field, while the even- and odd-parity modes relax on time scales set by $\varepsilon_\iota^{-1}$ and $\gamma_\lambda^{-1}$. The associated projected variables are defined as
\begin{subequations}
\label{eq:OP}
\begin{align}
T(\vec{r},t) &= T_{0} + \frac{1}{C} \sum_{\alpha} \left[n_{\alpha}-n^{\mathrm{eq}}_{\alpha}\right]\hbar\omega_{\alpha}, \label{eq:OP_local_T} \\
\theta_{\iota}(\vec{r},t) &= \sum_{\alpha} 2\sinh\!\left(\frac{\hbar\omega_{\alpha}}{2k_{\mathrm B}T_0}\right)\left[n_{\alpha}-n^{\mathrm{eq}}_{\alpha}\right]\nu_{\iota\alpha}, \\
\eta_{\lambda}(\vec{r},t) &= \sum_{\alpha} 2\sinh\!\left(\frac{\hbar\omega_{\alpha}}{2k_{\mathrm B}T_0}\right)\left[n_{\alpha}-n^{\mathrm{eq}}_{\alpha}\right]\chi_{\lambda\alpha}.
\end{align}
\end{subequations}
Equation~\eqref{eq:OP_local_T} is consistent with the local thermal-equilibrium construction summarized in Appendix~\ref{app:phonon}.

The odd-parity variables $\eta_\lambda$ represent flux-carrying modes and reconstruct the heat flux as
\begin{equation}
\label{eq:flux}
\vec{j}(\vec{r},t) = \frac{1}{V} \sum_{\lambda} \vec{\Lambda}_\lambda(\vec{r})\,\eta_{\lambda}(\vec{r},t),
\end{equation}
with coupling vectors
\begin{equation}
\label{eq:eigenmode_Lambda}
\vec{\Lambda}_\lambda = \sum_{\alpha} \frac{\chi_{\lambda\alpha}\hbar\omega_{\alpha}\vec{v}_{\alpha}}{2\sinh(\hbar\omega_{\alpha}/2k_{\mathrm B}T_0)}.
\end{equation}

In contrast, the even-parity variables $\theta_\iota$ carry no net energy flux and instead describe $\vec q$-symmetric distortions of the phonon distribution. These modes encode deviations from local equilibrium that persist even when $T(\vec r,t)\to T_0$, and therefore represent intrinsically nonlocal degrees of freedom.

\subsubsection{Memory and Nonlocality}
\label{sec:zwanzig_phBTE}

Expressed in terms of the projected variables, the dynamics become
\begin{subequations}
\label{eq:bte_r}
\begin{eqnarray}
C\,\partial_t T &=& - \sum_{\lambda} \vec{\Lambda}_\lambda \cdot \nabla \eta_{\lambda}, \\
\label{eq:bte_O_r}
\partial_t \eta_{\lambda} + \gamma_{\lambda}\eta_{\lambda} &=& - \frac{1}{k_{\mathrm B} T_0^2}\, \vec{\Lambda}_\lambda \cdot \nabla T - \sum_{\iota} \vec{\Pi}_{\lambda\iota} \cdot \nabla \theta_{\iota}, \\
\label{eq:bte_E_r}
\partial_t \theta_{\iota} + \varepsilon_{\iota}\theta_{\iota} &=& - \sum_{\lambda} \vec{\Pi}_{\lambda\iota} \cdot \nabla \eta_{\lambda}.
\end{eqnarray}
\end{subequations}

These equations provide a transparent decomposition of heat conduction into conserved, current-carrying, and nonlocal degrees of freedom. Temporal memory arises from the finite relaxation times $\gamma_\lambda^{-1}$ and $\varepsilon_\iota^{-1}$, while spatial nonlocality originates from the coupling between flux modes and nonlocal modes through
\begin{equation}
\label{eq:coupling_coefficients}
\vec{\Pi}_{\lambda\iota} = \sum_{\alpha} \chi_{\lambda\alpha}\nu_{\iota\alpha}\vec{v}_{\alpha}.
\end{equation}

This structure makes explicit that non-Fourier behavior is not imposed phenomenologically but emerges from the coupled dynamics of conserved energy, current-carrying modes, and nonlocal distribution distortions. It also provides the microscopic basis for the spatiotemporal kernel formulation developed in Sec.~\ref{sec:unified_framework}, in which the irrelevant variables are formally eliminated to yield a constitutive relation with intrinsic memory and nonlocality. The corresponding causal integral solutions, their inhomogeneous generalization, and the resulting hierarchy of memory- and nonlocality-induced corrections are developed in Appendix~\ref{app:homogeneous_phonon}--\ref{app:iterative_solution}. For convenience, Table~\ref{tab:zwanzig_variables} summarizes the projected variables, coupling quantities, and associated time scales introduced in this subsection.

Although formally equivalent to the conventional phBTE, the Zwanzig projection-operator provides a fundamentally different organization that makes the transport structure explicit. This structure enables systematic model reduction: modes whose relaxation times are short compared with the time scale of interest may be eliminated as instantaneous responses, while modes whose nonlocal coupling lengths are short compared with the spatial resolution yield local approximations.

\subsection{Reduced Constitutive Models}
\label{sec:reduced_models}
\begin{figure}[tb]
\includegraphics[width=0.75\columnwidth]{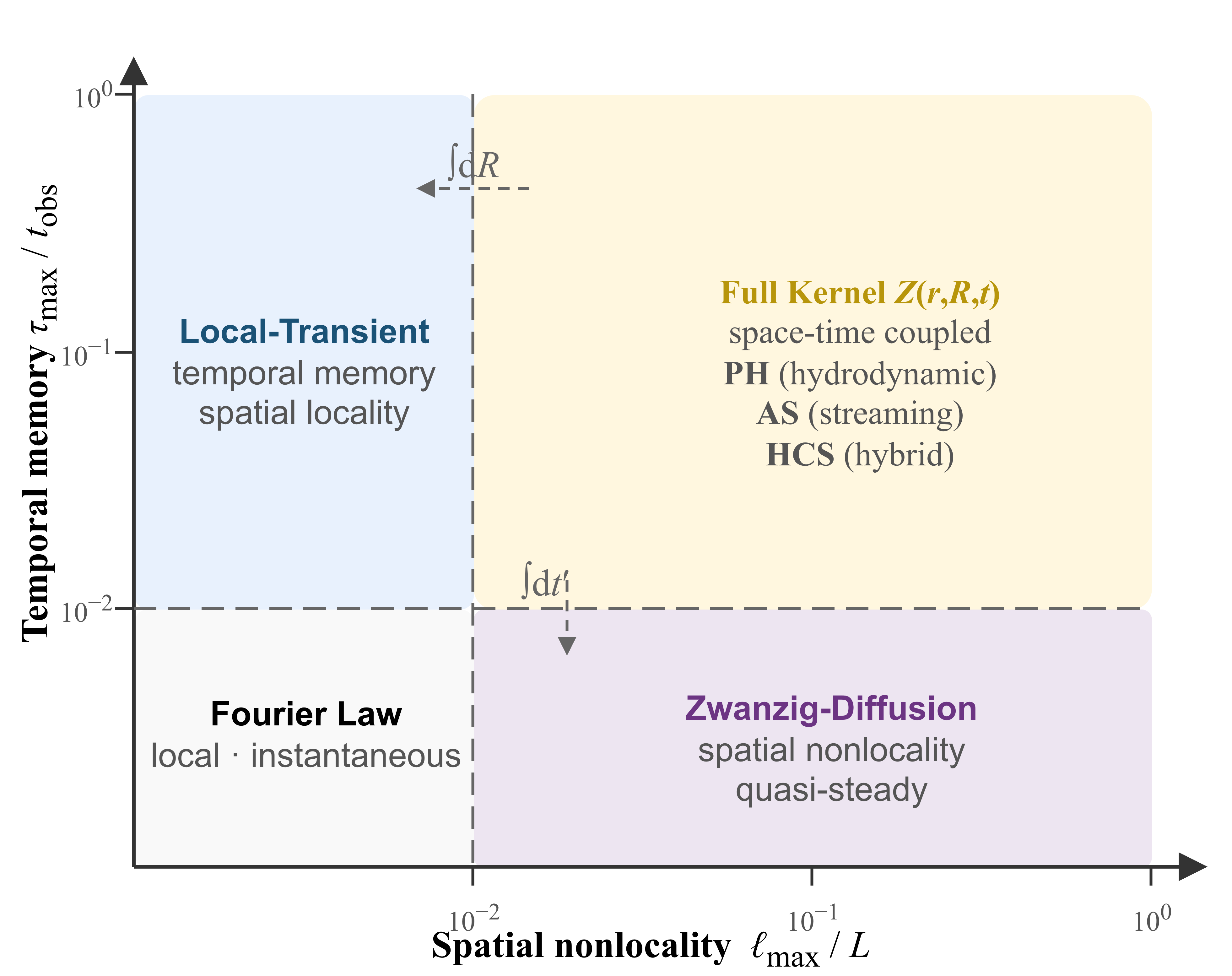}
\caption{Classification of heat-conduction regimes by temporal memory ($\tau_{\max}/t_{\mathrm{obs}}$, vertical axis) and spatial nonlocality ($\ell_{\max}/L$, horizontal axis), where $\tau_{\max}$ and $\ell_{\max}$ denote the longest phonon relaxation time and MFP, while $t_{\mathrm{obs}}$ and $L$ are the experimental time and length scales. When both ratios are small, Fourier's law applies (lower left). When both effects are comparable, the full spatiotemporal kernel is required, encompassing local--transient (LT), Zwanzig-diffusion (ZD), phonon-hydrodynamic (PH), attenuated-streaming (AS), and hybrid collective--streaming (HCS) regimes. Dashed lines labeled $\int d\vec{R}$ and $\int dt'$ indicate the reductions connecting the full kernel to the LT and ZD limits, respectively; their combined application recovers the Fourier conductivity $\kappa = \int dt'\!\int\!d\vec{R}\; Z$.}
\label{fig:regime_map}
\end{figure}

The utility of the Zwanzig-projected formulation lies not only in its formal rigor, but also in its ability to organize transport behavior into a hierarchy of reduced descriptions governed by scale separation. Rather than introducing separate constitutive models for different regimes, the present framework identifies them as controlled limits of a single underlying constitutive structure. The eigenvalue hierarchy and parity-based classification established in Sec.~\ref{sec:zwanzig} provide a principled basis for deriving reduced constitutive relations adapted to different transport regimes. Exploiting this structure, we identify four asymptotic limits of phonon-mediated heat conduction. Before deriving each regime in detail, we preview the overall classification in Fig.~\ref{fig:regime_map}; each region is developed in the subsections that follow. These limits are not intended to exhaust all possible transport behaviors, nor are they sharply exclusive. Rather, they represent physically transparent and mathematically controlled reductions of the underlying microscopic theory, in which selected dynamical processes remain leading while others become parametrically subleading. Technical details of the corresponding reductions are developed in Appendix~\ref{app:model-coupled-dynamics}.

The first two regimes arise when temporal memory and spatial nonlocality can be asymptotically separated. When spatial nonlocality is negligible but temporal memory remains significant, the transport reduces to the local--transient (LT) regime (Sec.~\ref{sec:ultrafast_and_local}), which is local in space but nonlocal in time. Conversely, when the evolution is slow compared with the intrinsic relaxation times while spatial correlations remain important, temporal memory may be averaged out, yielding the Zwanzig-diffusion (ZD) regime (Sec.~\ref{sec:quasi_steady}), which is quasi-steady in time but nonlocal in space.

When both temporal memory and spatial nonlocality remain asymptotically leading, transport cannot be reduced by eliminating either effect alone. In this regime, different structured limits arise depending on how the underlying modes are organized. In the phonon-hydrodynamic (PH) regime (Sec.~\ref{sec:phonon_hydrodynamics}), transport is governed by a small set of long-lived collective heat-flux modes selected by symmetry and conservation laws, while the remaining modes relax rapidly and can be systematically eliminated. In the attenuated-streaming (AS) regime (Sec.~\ref{sec:attenuated_streaming}), the dynamics are dominated by effectively paired odd- and even-parity modes, yielding a constitutive form that retains finite propagation together with temporal attenuation. Between these limits, a mixed regime can arise in which a subset of collective modes coexists with streaming contributions, reflecting incomplete separation of scales (Sec.~\ref{sec:hybrid_collective_streaming}).

We emphasize that temporal memory and spatial nonlocality generically coexist in the full constitutive kernel; the limits introduced below separate or reorganize these effects analytically to isolate the dominant physics at different scales. Taken together, these regimes organize a broad range of non-Fourier phenomena within a unified asymptotic hierarchy.

\subsubsection{Local--Transient Regime}
\label{sec:ultrafast_and_local}

We first consider the regime in which the coupling between flux modes and nonlocal modes is negligible, $\vec{\Pi}_{\lambda\iota}(\vec r)\to 0$, so that spatial nonlocality is suppressed while temporal memory remains essential. In this flux--nonlocal decoupled limit, the dynamics of the irrelevant variables reduce to
\begin{subequations}
\label{eq:bte_eigen_local}
\begin{align}
\label{eq:bte_O_local}
\frac{\partial \eta_{\lambda}(\vec r,t)}{\partial t} + \gamma_{\lambda}(\vec r)\,\eta_{\lambda}(\vec r,t) &\approx -\frac{1}{k_{\mathrm B}T_0^{2}}\, \vec{\Lambda}_{\lambda}(\vec r) \cdot \nabla T(\vec r,t),\\
\label{eq:bte_E_local} 
\theta_{\iota}(\vec r,t) &\approx 0.
\end{align}
\end{subequations}
The nonlocal-mode variables therefore do not contribute at leading order, and heat transport is governed entirely by local relaxation of the flux modes in response to the local temperature gradient.

Our previous work~\cite{crawford2024theory} analyzed this limit for an infinite homogeneous phonon gas. Extending that construction to finite and spatially inhomogeneous systems within a locally quasi-homogeneous approximation, we define the mode-resolved heat flux $\vec{j}_\lambda(\vec r,t)=V^{-1}\vec{\Lambda}_\lambda(\vec r)\,\eta_\lambda(\vec r,t)$ and the corresponding mode conductivity tensor $ \overleftrightarrow{\kappa}_\lambda(\vec r) = \frac{\vec{\Lambda}_\lambda(\vec r)\otimes\vec{\Lambda}_\lambda(\vec r)} {V k_{\mathrm B}T_0^2\gamma_\lambda(\vec r)}$. The constitutive relations in the local--transient regime then take the form
\begin{subequations}
\label{eq:constitutive_JT}
\begin{eqnarray}
\vec{j}(\vec{r},t) &=& \sum_{\lambda=1}^{M_\mathrm{O}} \vec{j}_\lambda(\vec{r},t), \label{eq:JT_mode_flux} \\
\bigl(1+\gamma^{-1}_\lambda(\vec r)\partial_t\bigr)\vec{j}_\lambda(\vec r,t) &=& -\,\overleftrightarrow{\kappa}_\lambda(\vec r)\cdot \nabla T(\vec r,t).
\label{eq:JT_j_lambda}
\end{eqnarray}
\end{subequations}
Here $\{\overleftrightarrow{\kappa}_\lambda(\vec r),\gamma_\lambda(\vec r)\}$ are the mode-resolved conductivity tensors and relaxation rates, and $M_\mathrm{O}$ is the number of odd-parity flux modes.

In practice, flux modes with comparable relaxation rates may be grouped into a smaller number of effective modes. This controlled coarse graining preserves the dominant memory structure while simplifying the constitutive description. The MCV equation corresponds to the limiting case in which all flux modes are represented by a single effective relaxation time. Equations~\eqref{eq:constitutive_JT} admit an equivalent causal kernel representation,
\begin{equation}
\label{eq:j_local_convolution}
\vec{j}(\vec{r},t) = -\int_{0}^{\infty} dt'\; \overleftrightarrow{Z}_\text{LT}(\vec r,t') \cdot \nabla T(\vec r,t-t'),
\end{equation}
where causality is enforced by the lower limit of integration. The local--transient kernel is
\begin{equation}
\label{eq:Z_t_r}
\overleftrightarrow{Z}_\text{LT}(\vec r,t) = \sum_{\lambda=1}^{M_\mathrm{O}} \gamma_\lambda(\vec r)\, e^{-\gamma_\lambda(\vec r)t}\, \overleftrightarrow{\kappa}_\lambda(\vec r)\, \Theta(t),
\end{equation}
where $\Theta(t)$ is the Heaviside step function enforcing causality. This is simply an exact rewriting of the mode-resolved constitutive dynamics in kernel form.

The temporal decay of $\overleftrightarrow{Z}_\text{LT}(\vec r,t)$ encodes intrinsic memory effects associated with ultrafast heat transport and provides a microscopic basis for local non-Fourier transient behavior, including regimes consistent with second-sound-like response~\cite{crawford2024theory}. In confined or spatially varying systems, the position dependence of $\gamma_\lambda(\vec r)$ and $\vec{\Lambda}_\lambda(\vec r)$ induces a corresponding position dependence of the LT kernel, providing a microscopic route to size- and geometry-dependent transport beyond bulk Fourier theory. Its time integral recovers the local conductivity tensor in the quasi-homogeneous limit, $ \int_0^\infty dt\;\overleftrightarrow{Z}_\text{LT}(\vec r,t) = \overleftrightarrow{\kappa}(\vec r) = \sum_{\lambda=1}^{M_\mathrm{O}} \overleftrightarrow{\kappa}_\lambda(\vec r)$. The corresponding causal integral form of the projected equations is derived in Appendix~\ref{app:homogeneous_phonon}.

\subsubsection{Zwanzig-Diffusion Regime}
\label{sec:quasi_steady}

We next consider the complementary regime in which temporal memory is negligible at the scale of observation, while spatial nonlocality remains essential. This corresponds to a slow-process or quasi-steady limit, commonly described as nonlocal diffusion. To distinguish it from conventional local diffusion governed by Fourier's law, we refer to this regime as \emph{Zwanzig diffusion} (ZD), a label that emphasizes its origin in Zwanzig's projection-operator reduction rather than in phenomenological nonlocal-diffusion models.

The ZD regime applies when the characteristic time scale of temperature evolution is much longer than the relaxation times of both the odd-parity flux modes, $\gamma_\lambda^{-1}$, and the even-parity nonlocal modes, $\varepsilon_\iota^{-1}$. Under these conditions, the explicit time dependence of the auxiliary variables may be neglected to leading order, while the spatial coupling between flux and nonlocal modes must still be retained. Heat transport is therefore quasi-steady in time but intrinsically nonlocal in space. In this controlled slow-process limit, Eqs.~\eqref{eq:bte_O_r} and \eqref{eq:bte_E_r} reduce to a closed equation for the flux-mode amplitudes,
\begin{equation}
\label{eq:bte_eta_only_r_ss}
\eta_{\lambda}(\vec r) =-\,\frac{\gamma_{\lambda}^{-1}(\vec r)}{k_{\mathrm B}T_0^2}\, \vec{\Lambda}_\lambda(\vec r)\cdot \nabla T(\vec r) + \gamma_{\lambda}^{-1}(\vec r) \sum_{\lambda'=1}^{M_{\mathrm O}} \sum_{\iota=1}^{M_{\mathrm E}} \vec{\Pi}_{\lambda\iota}(\vec r)\cdot \nabla \!\left[ \varepsilon_{\iota}^{-1}(\vec r)\, \vec{\Pi}_{\lambda'\iota}(\vec r)\cdot \nabla \eta_{\lambda'}(\vec r) \right],
\end{equation}
where $M_{\mathrm E}$ is the number of even-parity nonlocal modes. This equation may be solved iteratively as a systematic expansion in progressively higher-order spatial derivatives of the temperature field; details are given in Appendix~\ref{app:iterative_solution}. Substituting Eq.~\eqref{eq:bte_eta_only_r_ss} into the flux reconstruction formula~\eqref{eq:flux} and collecting terms by derivative order yields
\begin{equation}
\label{eq:j_ss_nonlocal_kappa}
\vec j(\vec r) = - \sum_{n_1,n_2,n_3=0}^{\infty} \frac{\overleftrightarrow{\kappa}^{(N)}(\vec r;n_1,n_2,n_3)} {n_1!n_2!n_3!} \cdot \nabla\!\left[ (\partial_x)^{n_1} (\partial_y)^{n_2} (\partial_z)^{n_3} T(\vec r) \right],
\end{equation}
where $N=n_1+n_2+n_3$. Equation~\eqref{eq:j_ss_nonlocal_kappa}, hereafter the \emph{Zwanzig-diffusion equation}, provides a systematic nonlocal generalization of Fourier's law. The zeroth-order tensor $\overleftrightarrow{\kappa}^{(0)}(\vec r;0,0,0)$ recovers the local conductivity, while the higher-order tensors encode progressively nonlocal corrections arising from coupling between flux modes and nonlocal modes. All conductivity tensors may be expressed explicitly in terms of the microscopic quantities $\gamma_\lambda(\vec r)$, $\varepsilon_\iota(\vec r)$, $\vec{\Lambda}_\lambda(\vec r)$, and $\vec{\Pi}_{\lambda\iota}(\vec r)$, so that the ZD regime provides a constitutive description of steady or quasi-steady nonlocal transport that remains microscopically anchored and, in principle, directly computable from first-principles calculations or atomistic simulations.

The infinite hierarchy in Eq.~\eqref{eq:j_ss_nonlocal_kappa} should be interpreted as an asymptotic gradient expansion rather than a uniformly convergent series. It is controlled when the temperature field varies smoothly over length scales large compared with the dominant nonlocal coupling lengths; when those scales become comparable, the expansion may converge slowly or fail, and one must revert to a fuller kernel representation. We note that the numerical illustrations in Sec.~\ref{sec:numerical_simulation} evaluate the full spatiotemporal kernel directly [Eq.~\eqref{eq:RTA_kernel_qw}] and therefore do not rely on convergence of this truncated gradient expansion.

An equivalent and often more compact formulation is obtained by resumming the gradient expansion into a spatial kernel. The starting point is the Taylor expansion
\begin{equation}
\label{eq:vector_taylor}
\nabla T(\vec R) = \nabla T(\vec r+\delta\vec r) = \sum_{n_1,n_2,n_3=0}^{\infty} \frac{(\delta x)^{n_1}(\delta y)^{n_2}(\delta z)^{n_3}} {n_1!\,n_2!\,n_3!}\, \nabla [\partial_x^{n_1}\partial_y^{n_2}\partial_z^{n_3} T(\vec r)],
\end{equation}
where $\delta\vec r= \vec R - \vec r =(\delta x,\delta y,\delta z)$. Identifying the coefficients in Eq.~\eqref{eq:j_ss_nonlocal_kappa} with the corresponding spatial moments then yields
\begin{equation}
\label{eq:J_ss_kernel_r_R}
\vec{j}(\vec{r}) = - \int_{\mathbb{V}} d^3\vec{R}\; \overleftrightarrow{Z}_\text{ZD}(\vec r,\vec R)\cdot \nabla T(\vec R),
\end{equation}
where $\overleftrightarrow{Z}_\text{ZD}(\vec r,\vec R)$ is the spatial Zwanzig-diffusion kernel. The generalized conductivity tensors are the spatial moments of this kernel,
\begin{equation}
\label{eq:general_kappa}
\overleftrightarrow{\kappa}^{(N)}(\vec r;n_1,n_2,n_3) = \int_{\mathbb V} d^3 \delta\vec r\ \overleftrightarrow{Z}_\text{ZD}(\vec r, \vec r + \delta \vec r)\, (\delta x)^{n_1}(\delta y)^{n_2}(\delta z)^{n_3}.
\end{equation}
For homogeneous bulk systems, translational invariance reduces $\overleftrightarrow{Z}_\text{ZD}(\vec r,\vec R)$ to a function of $\vec R-\vec r$. More importantly, without imposing translational invariance, the same constitutive form naturally accommodates finite-size effects and smoothly varying inhomogeneity. Boundary conditions may then be imposed directly at the kernel level, which is often more transparent than working within a truncated gradient expansion. In general, the ZD kernel does not admit a closed analytical form; explicit expressions are obtained below for the phonon-hydrodynamic and attenuated-streaming limits, in which additional scale separation constrains the mode structure.
%Sharp interfaces require additional treatment because the relevant phonon modes change qualitatively across the boundary; this motivates the Dyson-type kernel construction developed separately.

\subsubsection{Phonon-Hydrodynamic Regime}
\label{sec:phonon_hydrodynamics}

A fully general treatment of transient and nonlocal heat transport is analytically demanding. In the phonon-hydrodynamic (PH) regime, however, a clear separation of time scales enables a controlled reduction. When momentum-conserving scattering dominates over momentum-relaxing scattering, heat transport is governed by a small number of long-lived collective excitations, while all remaining microscopic modes relax rapidly. As shown in Appendix~\ref{app:GK}, retaining only the three slow odd-parity heat-flux modes and systematically eliminating the fast degrees of freedom yields a closed anisotropic hydrodynamic constitutive relation. We note that the same G--K form emerges whenever a small number of flux modes are well separated in relaxation rate from the remaining degrees of freedom~\cite{sendra2021derivation,guo2018phonon}; the label ''phonon hydrodynamic'' is retained to emphasize the physical context in which this separation is most pronounced (e.g., graphite and isotopically purified crystals at cryogenic temperatures~\cite{machida2018observation,huberman2019observation}), not to restrict the equation's domain of applicability.

In this regime, the heat flux is decomposed into three hydrodynamic components $\vec j_\lambda(\vec r,t)$, with $\lambda=1,2,3$, obeying
\begin{equation}
\label{eq:gk_like_general}
\Bigl(1+\gamma_\lambda^{-1}\partial_t\Bigr)\vec{j}_\lambda - \gamma_\lambda^{-1}\,\vec{\Lambda}_\lambda \sum_{\lambda'=1}^{3} \Biggl[ \vec{\beta}_{\lambda'}\!\cdot\!  \sum_{I,J}\mu_{\lambda\lambda' IJ}\,\partial^{2}_{IJ}\vec{j}_{\lambda'} \Biggr] = -\overleftrightarrow{\kappa}_\lambda \cdot \nabla T,
\end{equation}
where $\gamma_\lambda^{-1}$ are the relaxation times of the hydrodynamic modes, $\overleftrightarrow{\kappa}_\lambda$ are the corresponding conductivity tensors, and $\mu_{\lambda\lambda' IJ}$ is the heat-flux viscosity tensor generated by coupling to the eliminated fast modes. Crystal symmetry constrains $\mu_{\lambda\lambda' IJ}$ and thereby determines the anisotropic structure of the hydrodynamic response.

In the isotropic limit, Eq.~\eqref{eq:gk_like_general} reduces to the G--K equation,
\begin{equation}
\label{eq:gk_derived}
\Bigl( 1+\tau\partial_t -\tau\bigl(\mu_1 \nabla^2+\mu_2\, \nabla \nabla\!\cdot\bigr) \Bigr) \vec{j} = -\kappa_0\, \nabla T,
\end{equation}
thereby recovering its familiar form from a microscopic projection-based construction, without introducing a drift velocity field. Equivalent microscopic parameter expressions for the G--K equation have been obtained from moment-expansion approaches to the phBTE~\cite{sendra2021derivation,guo2018phonon,sendra2022hydrodynamic}; the present derivation embeds this equation within a unified kernel framework that simultaneously yields its spatiotemporal kernel form and its connections to other transport regimes. This three-mode truncation is exact in the sense that no further slow modes contribute when the time-scale separation is sharp. However, in materials with a broad distribution of phonon MFP, higher-order modes decay on comparable time scales and the G--K reduction becomes insufficient; the full kernel then captures transport features that no finite-order gradient expansion can represent. This limitation motivates the full spatiotemporal kernel developed in Sec.~\ref{sec:unified_framework}. The detailed hydrodynamic reduction and the symmetry structure of the associated nonlocal operator are given in Appendix~\ref{app:GK}.

Since only $\nabla\!\cdot\!\vec j$ enters the energy conservation law, one may restrict attention to curl-free heat fluxes. In that case,
\begin{equation}
\label{eq:gk_curl_free}
\Bigl(1+\tau\,\partial_t-\tau\mu \nabla^2\Bigr)\vec j = -\kappa_0\,\nabla T, \qquad \mu=\mu_1+\mu_2.
\end{equation}
This equation admits an equivalent spatiotemporal kernel representation (derived in Appendix~\ref{app:GK}),
\begin{equation}
\label{eq:gk_in_kernel}
\vec j(\vec r,t) = - \int_{0}^{\infty} dt' \int d^3\vec R\; \overleftrightarrow{\mathbb{Z}}_\text{PH}(\vec r,\vec R,t')\cdot \nabla T(\vec R,t-t'),
\end{equation}
with
\begin{equation}
\label{eq:GK_kernel}
\overleftrightarrow{\mathbb{Z}}_\text{PH} = \frac{\kappa_0}{\tau}\, \frac{e^{-t/\tau}}{(4\pi\mu t)^{3/2}} \exp\!\left(-\frac{|\vec R-\vec r|^2}{4\mu t}\right)\Theta(t).
\end{equation}
This kernel makes explicit the coexistence of exponential temporal memory and diffusive spatial spreading in the hydrodynamic regime. The PH kernel connects naturally to simpler limits. Spatial integration yields the local--transient kernel,
\begin{equation}
Z_{\mathrm{LT\text{-}PH}}(t)= \frac{\kappa_0}{\tau} e^{-t/\tau},
\end{equation}
which recovers the MCV form. Conversely, time integration yields the spatial diffusion kernel,
\begin{equation}
Z_{\text{ZD-PH}}(\vec r,\vec R) = \frac{\kappa_0}{4\pi \ell^2}\, \frac{e^{-|\vec R-\vec r|/\ell}}{|\vec R-\vec r|}, \qquad \ell=\sqrt{\tau\mu},
\end{equation}
corresponding to a Yukawa-type nonlocal diffusion law. Thus both local transient transport and steady nonlocal diffusion appear as controlled reductions of the hydrodynamic kernel.

\subsubsection{Attenuated-Streaming Regime}
\label{sec:attenuated_streaming}

When the collective coupling is weak, heat transport is dominated by finite-speed propagation  between scattering events rather than collective diffusion. In this limit, thermal disturbances propagate approximately ballistically along characteristic directions, while dissipation enters through temporal attenuation. This is closely related to the commonly discussed quasi-ballistic regime; we use the term \emph{attenuated streaming} (AS) to emphasize the coexistence of directional propagation and exponential decay.

A controlled reduction is obtained by retaining only mode-diagonal couplings between each odd-parity flux mode and its associated even-parity partner (Appendix~\ref{app:AS}). This paired structure captures the leading interplay between directional transport and temporal memory while neglecting higher-order intermode mixing. The resulting constitutive relation for each modal heat flux $\vec j_\lambda$ takes the factorized form
\begin{equation}
\label{eq:AS_constitutive_full}
\bigl(1+\tau_\lambda\partial_t+\vec{\ell}_\lambda\cdot\nabla\bigr) \bigl(1+\tau_\lambda\partial_t-\vec{\ell}_\lambda\cdot\nabla\bigr)\, \vec j_\lambda = -\,\overleftrightarrow{\kappa}_\lambda\cdot\nabla\bigl(T+\tau_\lambda\partial_t T\bigr),
\end{equation}
where $\tau_\lambda$ is the relaxation time and $\vec{\ell}_\lambda$ defines the characteristic streaming length and direction.

The factorized structure admits an immediate interpretation in terms of first-order attenuated-streaming operators $1+\tau_\lambda\partial_t \pm \vec{\ell}_\lambda\cdot\nabla$, which describe propagation along straight-line characteristics with finite velocity and exponential temporal decay. For a homogeneous medium, the AS regime is equivalently represented by the spatiotemporal kernel
\begin{equation}
\label{eq:AS_kernel}
\overleftrightarrow{\mathbb{Z}}_{\mathrm{AS}}(\delta \vec r,t) = \sum_{\lambda} \frac{\overleftrightarrow{\kappa}_\lambda}{2\tau_\lambda} e^{-t/\tau_\lambda} \Big[ \delta^{(3)}(\delta \vec r - \vec{\ell}_\lambda t/\tau_\lambda) + \delta^{(3)}(\delta \vec r + \vec{\ell}_\lambda t/\tau_\lambda) \Big]\Theta(t),
\end{equation}
which makes explicit the coexistence of finite-speed propagation and temporal attenuation.

In the local limit, spatial integration recovers a purely temporal memory kernel. In the steady limit, time integration yields a spatially nonlocal diffusion kernel describing exponentially attenuated transport along the streaming direction. As shown in Appendix~\ref{app:AS}, the attenuated-streaming regime contains the RTA as a limiting case, in which each phonon mode propagates ballistically with its group velocity between scattering events. The AS formulation therefore provides a constitutive bridge between microscopic streaming dynamics and continuum nonlocal transport models in the quasi-ballistic regime. The relation of this kernel-based formulation to the analytical Green-function solution of Hua and Minnich~\cite{hua2014analytical} is discussed in Appendix~\ref{app:MH_relation}.

\subsubsection{Hybrid Collective--Streaming Regime}
\label{sec:hybrid_collective_streaming}

Between the phonon-hydrodynamic and attenuated-streaming limits lies an intermediate regime in which only a subset of modes exhibits collective behavior, while the remainder contribute through quasi-ballistic streaming. This situation is generic in materials with broad distributions of relaxation times and MFP, where complete time-scale separation does not hold.

To capture this structure, we partition the projected variables into a slow collective subset $S$ and a complementary fast subset $F$. Within the slow subset, the full odd--even coupling is retained, yielding a reduced hydrodynamic description. The remaining modes are treated in the paired-coupling approximation introduced in Sec.~\ref{sec:attenuated_streaming}. Cross-couplings between the two sectors are subleading and neglected at leading order. Under this reduction, the total heat flux decomposes as $
\vec j = \vec j^{(S)} + \vec j^{(F)}$.

The collective component obeys a reduced hydrodynamic equation,
\begin{equation}
\label{eq:hybrid_collective_main}
\Bigl(1+\tau^{(S)}\partial_t - \tau^{(S)}\mathcal{D}^{(S)}\Bigr)\vec j^{(S)} = -\,\overleftrightarrow{\kappa}^{(S)}\cdot\nabla T,
\end{equation}
obtained by retaining full odd--even coupling within the slow block and eliminating the corresponding even-parity modes to leading order (Appendix~\ref{app:hybrid}).

The fast component is expressed as a superposition of attenuated-streaming channels,
\begin{equation}
\vec j^{(F)}=\sum_{\lambda\in F}\bigl(\vec j_{\lambda+}+\vec j_{\lambda-}\bigr),
\end{equation}
with the directional contributions satisfying
\begin{subequations}
\label{eq:hybrid_AS_main}
\begin{align}
\bigl(1+\tau_\lambda\partial_t+\vec{\ell}_\lambda\cdot\nabla\bigr)\vec j_{\lambda+} &= -\frac{1}{2}\overleftrightarrow{\kappa}_\lambda\cdot\nabla T, \\
\bigl(1+\tau_\lambda\partial_t-\vec{\ell}_\lambda\cdot\nabla\bigr)\vec j_{\lambda-} &= -\frac{1}{2}\overleftrightarrow{\kappa}_\lambda\cdot\nabla T.
\end{align}
\end{subequations}
This decomposition separates collective diffusive–hydrodynamic transport at long wavelengths from directional streaming at shorter scales. It reduces to the PH limit when the collective subset dominates and to the AS limit when collective coupling becomes negligible. The corresponding spatiotemporal kernel representation is given in Appendix~\ref{app:hybrid}.

\section{Unified Spatiotemporal Framework for Heat Conduction}
\label{sec:unified_framework}

Section~\ref{sec:phonon_conduction} established two central points. First, temporal memory and spatial nonlocality arise microscopically from the dynamics of energy transport, while familiar bulk transport coefficients emerge only after coarse graining over time and length scales large compared with the intrinsic relaxation times and nonlocal correlation lengths. Second, these effects need not be introduced through \emph{ad hoc} relaxation times or nonlocal lengths: within linear response, they are naturally encoded in a single spatiotemporal constitutive kernel. Motivated by these observations, we formulate heat conduction in a general material medium, crystalline, amorphous, composite, or spatially heterogeneous, in terms of the causal two-point kernel $\overleftrightarrow{\mathbb{Z}}(\vec r,\vec R,t\ge 0)$ introduced below in Eq.~\eqref{eq:j_convolution}. This viewpoint is not restricted to a phonon-gas description. In Sec.~\ref{sec:microscopic_foundation}, we show that $\overleftrightarrow{\mathbb{Z}}$ admits a rigorous microscopic definition as a space-projected equilibrium heat-flux time-correlation function within a spatiotemporal Green--Kubo framework, thereby identifying the kernel as a primary transport descriptor for nonuniform media under linear response, provided that an operational temperature field remains well defined at the coarse-graining scale probed by the kernel.

Like any formally exact constitutive object, the kernel $\overleftrightarrow{\mathbb{Z}}$ admits a hierarchy of controlled approximations, each producing a well-defined kernel whose asymptotic reductions, constitutive relations, and transport predictions follow from the same architecture. The RTA provides a complete, analytically tractable realization computable from standard first-principles phonon properties, capturing the full interplay of temporal memory and spatial nonlocality through the modal distribution of lifetimes and MFP. Hydrodynamic truncations retain a small number of collective modes and are appropriate when those modes dominate transport. For materials in which intermode coupling is significant, incorporation of the off-diagonal mode-coupling tensors $\vec{\Pi}_{\lambda\iota}$ recovers intermode correlations within the same formalism. Conventional transport parameters such as thermal conductivity and interfacial resistance arise as limiting projections of the kernel, while deviations from those limits encode the memory and nonlocal effects that become directly testable in ultrafast and nanoscale experiments.

\subsection{Unified Kernel-Based Constitutive Framework}
\label{sec:phenomenological_generalization}

We first define the causal two-point kernel and its associated evolution equation for general media (Sec.~\ref{sec:kernel}), then extend the framework to sharp material interfaces through a generalized Kapitza model (Sec.~\ref{sec:generalized_kapitza}).

\subsubsection{A Causal Two-Point Kernel}
\label{sec:kernel}

We write the unified constitutive relation as
\begin{equation}
\label{eq:j_convolution}
\vec{j}(\vec r,t) = - \int_{0}^{\infty} dt' \int_{\mathbb V} d^3\vec R\; \overleftrightarrow{\mathbb{Z}}(\vec r,\vec R,t') \cdot \nabla T(\vec R,t-t'),
\end{equation}
which is nonlocal in both space and time and, in general, inhomogeneous through its two spatial arguments. Causality is enforced by restricting the memory integral to $t'\ge 0$. Combining Eq.~\eqref{eq:j_convolution} with the local energy conservation law, $\rho(\vec r)c(\vec r)\,\partial_t T(\vec r,t)+\nabla\!\cdot\!\vec j(\vec r,t)=Q(\vec r,t)$, yields a closed integro-differential evolution equation for the temperature field,
\begin{equation}
\label{eq:T_dynamics}
\partial_t T(\vec r,t) - \int_{0}^{\infty} dt' \int_{\mathbb V} d^3\vec R\; \nabla \cdot \Bigg[ \frac{\overleftrightarrow{\mathbb{Z}}(\vec r,\vec R,t')}{\rho(\vec r)c(\vec r)} \cdot \nabla T(\vec R,t-t') \Bigg] = \frac{Q(\vec r,t)}{\rho(\vec r)c(\vec r)},
\end{equation}
where $Q(\vec r,t)$ is an externally imposed volumetric heating rate, set to zero when absent. Thus, $\overleftrightarrow{\mathbb{Z}}$ encodes temporal memory, spatial correlations, and material inhomogeneity within a single constitutive operator, while external forcing enters only through the conservation law.

Equation~\eqref{eq:j_convolution} presupposes a meaningful temperature field $T(\vec r,t)$ at the spatial resolution probed by the kernel. Within the projection-operator construction of Sec.~\ref{sec:phonon_conduction}, $T(\vec r,t)$ is the relevant variable proportional to the coarse-grained local energy density and remains well defined so long as projection onto this field is well posed. This requirement is weaker than strict local thermodynamic equilibrium, but it excludes the strictly collisionless limit, in which a scalar temperature ceases to be an adequate descriptor and a phase-space description becomes necessary.

The two-point structure in Eq.~\eqref{eq:j_convolution} naturally accommodates smoothly varying inhomogeneity, such as graded composition or distributed disorder, as well as finite-size effects within a single constitutive framework. Abrupt interfaces are qualitatively different: the energy-carrying states on either side of a sharp boundary need not share a common mode basis, and interfacial transmission and reflection cannot, in general, be captured by a local modification of the bulk kernel. Instead, the interface is represented by a spatially localized coupling that connects the bulk responses on either side, giving rise to an effective interfacial kernel that encodes mode conversion, transmission, and reflection, as developed in Sec.~\ref{sec:generalized_kapitza}.

It is worth distinguishing Eq.~\eqref{eq:j_convolution} from source-driven Green's-function solutions of the phBTE~\cite{hua2014analytical,hua2020space,chiloyan2021green}, in which one solves the linearized kinetic equation for a prescribed heating distribution and obtains the temperature response. Such Green functions are \emph{solution-level} objects tied to a particular closure and geometry. The present formalism instead emphasizes the \emph{constitutive-level} kernel $\overleftrightarrow{\mathbb{Z}}$, which specifies the intrinsic causal flux-force relation independent of the external heating history. Once $\overleftrightarrow{\mathbb{Z}}$ is known, the corresponding temperature Green function follows by operator inversion; for a homogeneous medium, $G_T(\vec k,\omega) = [i\omega\,\rho c + \vec k\!\cdot\!\overleftrightarrow{\mathbb{Z}}(\vec k,\omega)\!\cdot\!\vec k]^{-1}$, where $\vec k$ and $\omega$ are the spatial wavevector and temporal frequency, respectively. By making the separation between intrinsic material response and source-dependent contributions explicit, the kernel formulation provides a transferable descriptor that couples naturally to continuum solvers and admits systematic extension to heterogeneous media.

%%%%%%%%%%%%%%%%%%%%%%%%%%%%%%%%%%%%%%%%%%%%%%%%%%%%%%%%%%
\subsubsection{Generalized Kapitza Interface Model}
\label{sec:generalized_kapitza}
%%%%%%%%%%%%%%%%%%%%%%%%%%%%%%%%%%%%%%%%%%%%%%%%%%%%%%%%%%

The two-point kernel formalism is naturally suited to heat conduction in nonuniform media, where spatial heterogeneity and nonlocal correlations play an essential role. As a first example, we consider transport across a material interface. Thermal boundary resistance has been extensively studied since Kapitza's original observation at solid--helium interfaces~\cite{pollack1969kapitza}, with acoustic and diffuse mismatch models providing the standard theoretical framework for solid--solid interfaces~\cite{swartz1989thermal}; Chen \emph{et al.}~\cite{chen2022interfacial} provide a comprehensive recent review. Non-Markovian memory effects in interfacial heat exchange have been captured through a generalized Langevin equation approach~\cite{Zeng2021ITC}, though within a spatially lumped description. Within the present framework, interfacial resistance is not introduced as an external boundary condition, but arises from the spatial structure of the kernel itself. In the long-wavelength limit, this response can be coarse-grained into an effective lumped resistance. More generally, when the interface has finite spatial extent or when nonlocal correlations extend across it, the response retains spatial structure that cannot be reduced to a single scalar $R_K$.

%%%%%%%%%%%%%%%%%%%%%%%%%%%%%%%%%%%%%%%%%%%%%%%%%%%%%%%%%%
\begin{table}[tb]
\caption{
Parameters used in the finite-rank interfacial kernel model  for the results shown in Figs.~\ref{fig:kapitza_interface} and \ref{fig:kapitza_Tprofile}. The parameters $\kappa_{A,B}$ and $\ell_{A,B}$ define the bulk nonlocal kernels in each medium, $\xi$ sets the interfacial width, and $\mathcal{C}_{mm'}$ controls the strength of intra- and cross-interface coupling.
}
\label{tab:kapitza_params}
\centering
\begin{tabular}{c c}
\hline\hline
Parameter & Value \\
\hline
$\kappa_A$ & 130 W\,m$^{-1}$\,K$^{-1}$ \\
$\kappa_B$ & 60  W\,m$^{-1}$\,K$^{-1}$ \\
$\ell_A$ & 65 nm \\
$\ell_B$ & 25 nm \\
$\xi$    & 1.7 nm \\
$\mathcal{C}_{AA}$ & $-0.537$ nm$^2$\,K/W \\
$\mathcal{C}_{BB}$ & $-0.498$ nm$^2$\,K/W \\
$\mathcal{C}_{AB}=\mathcal{C}_{BA}$ & $0.107$ nm$^2$\,K/W \\
\hline
$j_0$            & 8.32 GW\,m$^{-2}$ \\
$R_K$            & 3.25 nm$^2$\,K/W\\
\hline\hline
\end{tabular}
\end{table}
%%%%%%%%%%%%%%%%%%%%%%%%%%%%%%%%%%%%%%%%%%%%%%%%%%%%%%%%%%

To make this connection explicit, consider steady one-dimensional transport across an interface at \(x=0\), with semi-infinite material \(A\) of bulk conductivity \(\kappa_A\) in \(x<0\) and semi-infinite material \(B\) of bulk conductivity \(\kappa_B\) in \(x>0\). In the steady limit, the full spatiotemporal kernel reduces to the Zwanzig-diffusion kernel \(Z(x,X)\). For a constant heat flux \(j_0\), we write the steady-state temperature gradient as
\begin{equation}
\label{eq:grad_decomp}
\frac{dT_{ss}(x)}{dx} = -\Theta(-x)\frac{j_0}{\kappa_A} -\Theta(x)\frac{j_0}{\kappa_B} -j_0 s(x),
\end{equation}
where \(s(x)\) is a localized boundary-layer correction satisfying \(s(x)\to 0\) as \(|x|\to\infty\). Using the steady constitutive relation, $ j_0=-\int_{-\infty}^{\infty} dX\, Z(x,X)\,\frac{dT_{ss}(X)}{dX}$, one obtains the integral equation
\begin{equation}
\label{eq:f_integral_equation_split}
\int_{-\infty}^{\infty} dX\, Z(x,X)\, s(X) = 1 -\frac{1}{\kappa_A}\int_{-\infty}^{0} dX\, Z(x,X) -\frac{1}{\kappa_B}\int_{0}^{\infty} dX\, Z(x,X).
\end{equation}

\begin{figure}[!htbp]
\centering
\includegraphics[width=3in]{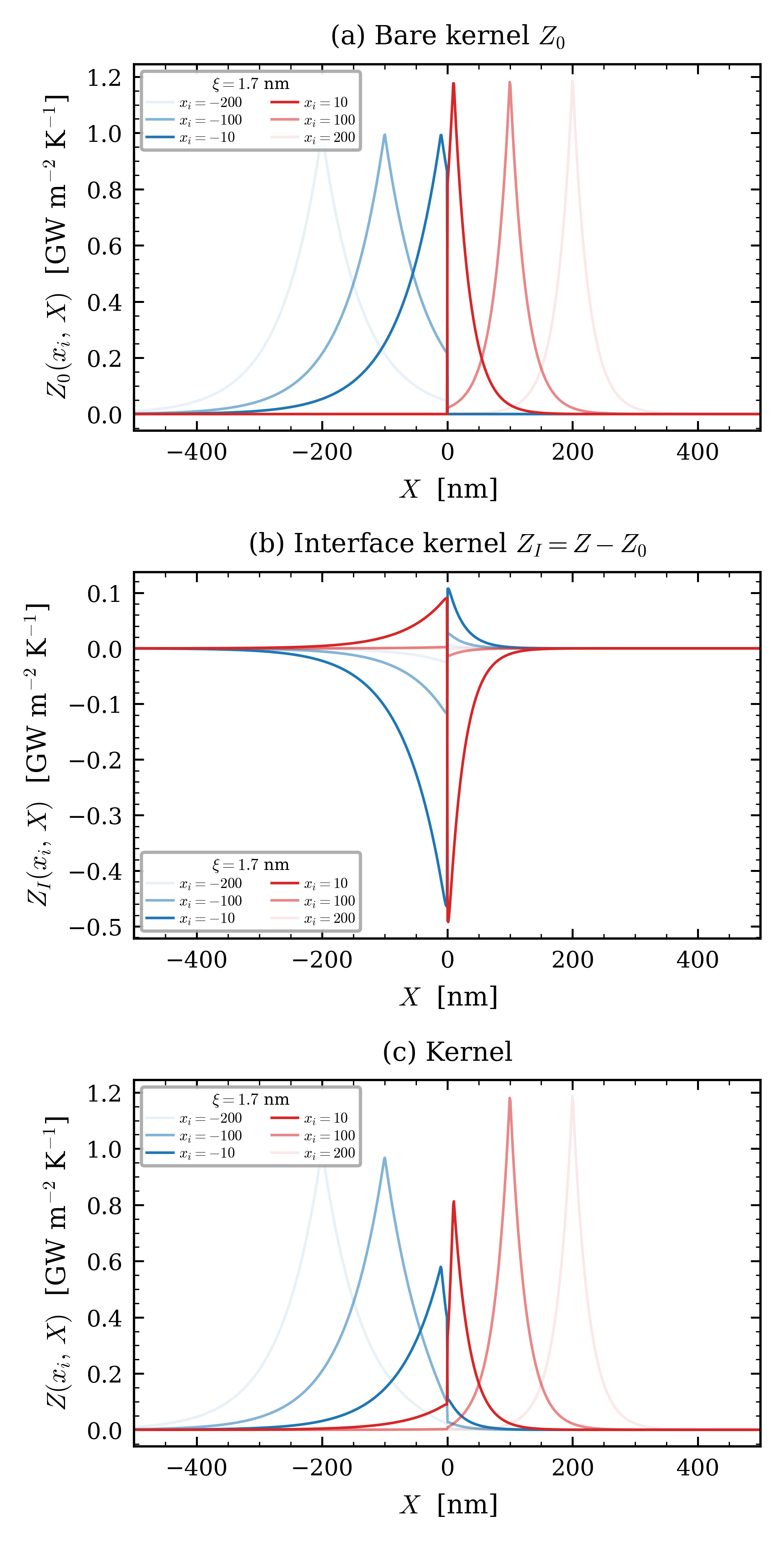}
\caption{Kernel decomposition for a model 1D interface.  (a)~Bare block-diagonal kernel $Z_0(x_i,X)$ for representative source points $x_i$ in materials A ($x_i<0$, blue) and B ($x_i>0$, red), showing exponential nonlocal spreading within each medium and no cross-interface coupling.  (b)~Interfacial kernel $Z_I(x_i,X)=Z-Z_0$, which is spatially localized near $x=0$ and introduces cross-interface correlations with opposite-signed contributions on either side of the interface. (c)~Full dressed kernel $Z(x_i,X)=Z_0+Z_I$, exhibiting redistribution of weight near the interface while preserving bulk behavior far from $x=0$.}
\label{fig:kapitza_interface}
\end{figure}

\begin{figure}[!htbp]
\centering
\includegraphics[width=3in]{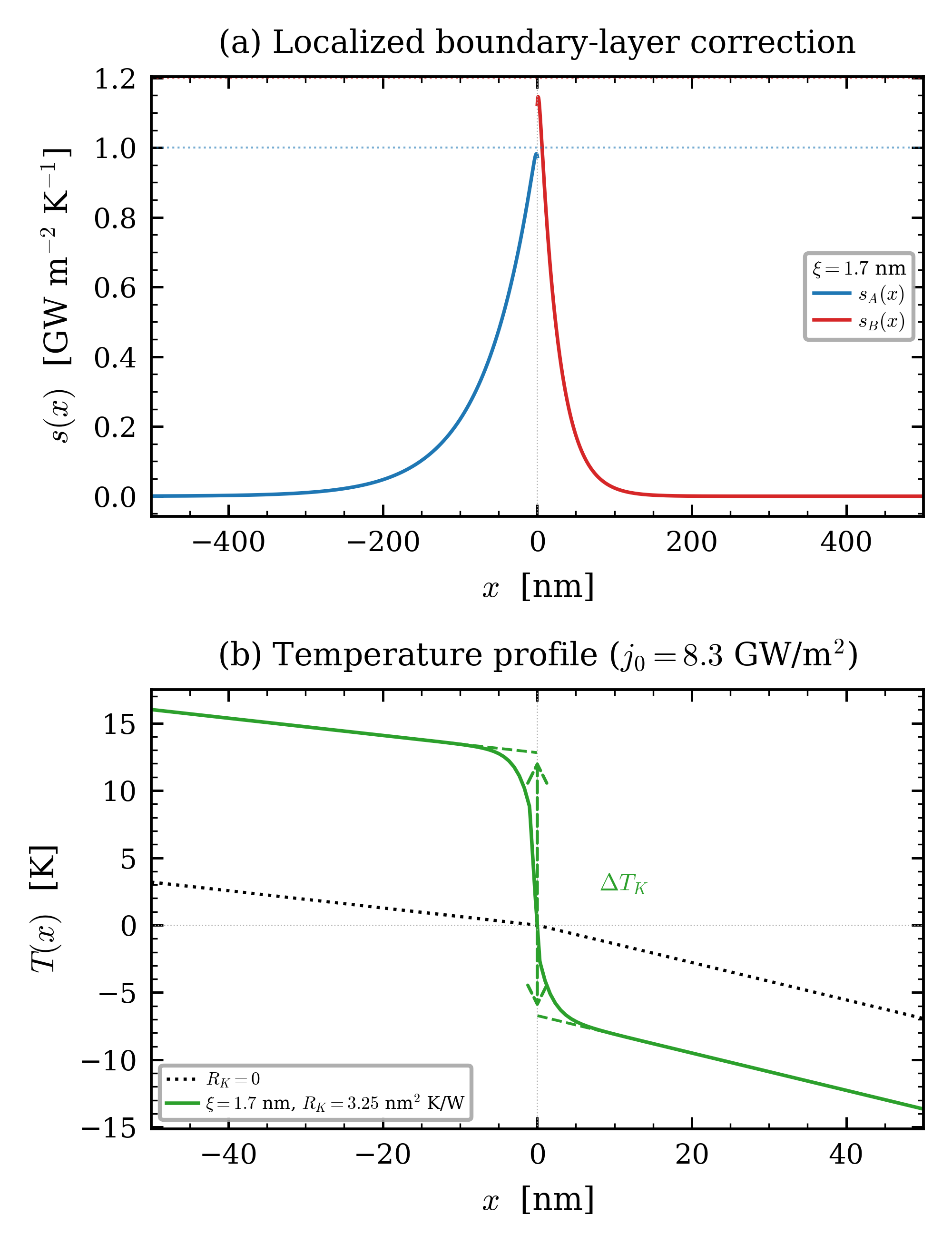}
\caption{Boundary-layer structure and steady-state temperature profile for the model 1D interface of Fig.~\ref{fig:kapitza_interface}. (a)~Boundary-layer correction $s(x)$ [Eq.~\eqref{eq:grad_decomp}], whose integral gives the effective Kapitza resistance $R_K=\int s(x)\,dx$ [Eq.~\eqref{eq:DeltaTK_revised}]. The localized peak near $x=0$ reflects the interfacial distortion of the temperature gradient. (b)~Steady-state temperature profile $T_{ss}(x)$ under constant heat flux $j_0=8.32~\mathrm{GW/m^2}$. The dotted line shows the no-interface ($R_K=0$) reference, while the solid curve corresponds to the dressed kernel with finite interfacial width $\xi=1.7~\mathrm{nm}$, yielding an effective Kapitza resistance $R_K=3.25~\mathrm{nm^2\,K/W}$.}
\label{fig:kapitza_Tprofile}
\end{figure}

Once the Zwanzig-diffusion kernel in the interfacial region is specified, Eqs.~\eqref{eq:grad_decomp} and \eqref{eq:f_integral_equation_split} determine the steady temperature profile \(T_{ss}(x)\). Within this description, the temperature field remains continuous; the commonly defined temperature jump is therefore not a literal discontinuity, but a reduced quantity associated with the interfacial boundary layer. A coordinate-independent definition is
\begin{equation}
\label{eq:DeltaTK_revised}
R_K=\frac{\Delta T_K}{j_0} = \int_{-\infty}^{+\infty} s(x)\,dx,
\end{equation}
where $\Delta T_K$ is the apparent temperature jump across the interface, which measures the excess temperature drop relative to the extrapolated bulk linear profiles. In this sense, the conventional Kapitza resistance is reinterpreted as a coarse-grained descriptor of the underlying interfacial response encoded in the kernel.

The central difficulty is the construction of a physically meaningful interfacial kernel \(Z(x,X)\). Here, we adopt a finite-rank separable model,
\begin{equation}
\label{eq:Z_dress}
Z_{\mathrm{ZD}}(x,X)=Z_0(x,X)+Z_I(x,X),
\end{equation}
where the block-diagonal reference kernel is
\begin{equation}
\label{eq:Z0_block}
Z_0(x,X) = \Theta(-x)\Theta(-X)\,\frac{\kappa_A}{2\ell_A}e^{-|X-x|/\ell_A} + \Theta(x)\Theta(X)\,\frac{\kappa_B}{2\ell_B}e^{-|X-x|/\ell_B},
\end{equation}
with $\ell_{A,B}$ the characteristic nonlocal lengths of materials $A$ and $B$, and the interfacial contribution is taken as $Z_I(x,X) = \sum_{m,m'\in\{A,B\}} w_m(x)\,\mathcal{C}_{mm'}\,w_{m'}(X)$,with localized basis functions $w_A(x)=\Theta(-x)e^{x/\xi}$ and $w_B(x)=\Theta(x)e^{-x/\xi}$. Here $\xi$ sets the interfacial width and $\mathcal{C}_{mm'}$ controls the strength of intra- and cross-interface coupling.

The present construction is not intended as a quantitative interface model, but as a demonstration of how interfacial resistance emerges structurally within the kernel formalism. In particular, the separable form provides a minimal representation of cross-interface correlations and their influence on transport, rather than a unique or material-specific description.

Far from the interface, \(Z_I\to 0\), implying \(s(x)\to 0\) and recovery of the bulk gradients \(dT_{ss}/dx=-j_0/\kappa_{A,B}\). Near \(x=0\), however, the gradient is modified over a boundary layer set by the spatial extent of \(Z_{\mathrm{ZD}}(x,X)\). The parameters used in constructing the model kernel and temperature profile in Figs.~\ref{fig:kapitza_interface} and \ref{fig:kapitza_Tprofile} are summarized in Table~\ref{tab:kapitza_params}; the effective Kapitza resistance $R_K$ is obtained from the steady-state profile via Eqs.~\eqref{eq:grad_decomp} and \eqref{eq:DeltaTK_revised}. The parameters are chosen to illustrate the structural interpretation by which Kapitza resistance arises within the kernel framework, rather than to represent a specific material interface.

Within this picture, interfacial resistance can be understood as arising from a redistribution of spatial correlations. The bare kernel $Z_0$ [Fig.~\ref{fig:kapitza_interface}(a)] confines transport within each medium, with the characteristic nonlocal lengths $\ell_{A,B}$ setting the spatial extent. The interfacial contribution $Z_I$ [Fig.~\ref{fig:kapitza_interface}(b)] introduces cross-interface correlations while reducing same-side contributions, reflecting transmission and reflection processes at the boundary in an averaged sense. The resulting dressed kernel [Fig.~\ref{fig:kapitza_interface}(c)] retains bulk behavior far from the interface while exhibiting significant modification within a boundary layer of width $\sim\xi$.

The boundary-layer correction $s(x)$, shown in Fig.~\ref{fig:kapitza_Tprofile}(a), is localized near the interface where the kernel redistribution is strongest; its integral directly yields $R_K$ [Eq.~\eqref{eq:DeltaTK_revised}]. The corresponding steady-state temperature profile [Fig.~\ref{fig:kapitza_Tprofile}(b)] shows that the apparent temperature jump arises from a localized distortion of the gradient rather than a discontinuity in $T(x)$, consistent with Eq.~\eqref{eq:grad_decomp}. These results illustrate how interfacial thermal resistance can be interpreted as a consequence of the spatial structure of the kernel, providing a bridge between nonlocal constitutive descriptions and conventional boundary resistance models.

\subsection{Microscopic Foundation}
\label{sec:microscopic_foundation}

The continuum kernel introduced above admits a rigorous microscopic definition. We first establish a spatiotemporal Green--Kubo relation that identifies the kernel as a space-resolved heat-flux correlation function (Sec.~\ref{sec:spatiotemporal_GK}), then introduce the bond-centered energy-flux operator needed to evaluate that correlation from atomistic trajectories (Sec.~\ref{sec:Hardy_flux}).

\subsubsection{Spatiotemporal Green--Kubo Relation}
\label{sec:spatiotemporal_GK}

Within linear response, irreversible transport is determined by equilibrium fluctuations: microscopic current correlations define macroscopic kinetic coefficients through Green--Kubo relations. For heat conduction in a macroscopically homogeneous medium, the conventional Green--Kubo formula expresses the bulk thermal conductivity tensor as the time integral of the autocorrelation of the \emph{system heat current}, or equivalently the spatially averaged heat flux. This bulk expression is therefore already a coarse-grained object, obtained by integrating out spatial structure and retaining only the long-wavelength ($k\!\to\!0$) response. To emphasize this point in notation consistent with the present framework, we write the local--transient bulk kernel as $\overleftrightarrow{Z}^{\mathrm{bulk}}_\text{LT}(t) = \frac{V}{k_B T_0^2} \left\langle \vec{j}_{\mathrm{bulk}}(t)\otimes \vec{j}_{\mathrm{bulk}}(0) \right\rangle_{\mathrm{eq}}$,
where $\vec{j}_{\mathrm{bulk}}(t) = V^{-1}\int_{\mathbb V} d^3\vec r\, \vec{j}(\vec r,t)$ and $\vec{J}(t) = \int_{\mathbb V} d^3\vec r\, \vec{j}(\vec r,t) = V\vec{j}_{\mathrm{bulk}}(t)$. Time integration of the L--T bulk kernel recovers the bulk conductivity tensor $\overleftrightarrow{\kappa}$.

The equilibrium ensemble, however, already contains the full spatial structure through the fluctuating microscopic energy-flux density $\vec{j}(\vec r,t)$. Retaining that resolution leads to a \emph{spatiotemporal} Green--Kubo relation,
\begin{equation}
\label{eq:gGK_full}
\overleftrightarrow{\mathbb Z}(\vec r,\vec R,t) = \frac{1}{k_B T_0^2} \left\langle \vec j(\vec r,t)\otimes \vec j(\vec R,0) \right\rangle_{\mathrm{eq}}, \qquad (t>0),
\end{equation}
which we take as the fundamental constitutive descriptor of heat conduction under linear response. Equation~\eqref{eq:gGK_full} generalizes the conventional Green--Kubo construction structurally rather than procedurally: the usual bulk coefficient is recovered only after integrating the kernel over space and time. In particular, inserting $\vec{J}(t)=\int d^3\vec r\,\vec j(\vec r,t)$ shows directly that the standard bulk Green--Kubo relation follows from integrating Eq.~\eqref{eq:gGK_full} over both spatial arguments.

Crucially, Eq.~\eqref{eq:gGK_full} is formulated directly in terms of microscopic energy-flux operators and does not rely on a quasiparticle or phonon-gas representation. It is therefore applicable, in principle, to crystalline, amorphous, composite, and spatially heterogeneous systems, provided that (i) a temperature field is operationally well defined at the kernel’s coarse-graining scale and (ii) the dynamics can be linearized about global equilibrium. Within linear response, Eq.~\eqref{eq:gGK_full} defines the kernel uniquely: for a given Hamiltonian and reference equilibrium state, the two-point heat-flux correlation function is fully determined, with no residual closure ambiguity. A rigorous derivation of Eq.~\eqref{eq:gGK_full} from a time-dependent projection formalism assuming local thermal equilibrium and linear response is given in Appendix~\ref{app:Robertson_Z}. The projection-operator construction underlying the spatiotemporal kernel was originally developed by Robertson~\cite{robertson1966,robertson1967} using a generalized canonical density operator; the present work recasts that formalism into an explicit constitutive kernel suitable for continuum coupling in nonuniform media.

In practice, evaluating Eq.~\eqref{eq:gGK_full} extends standard Green--Kubo calculations by requiring two-point correlations, $\langle \vec j(\vec r,t) \otimes \vec j(\vec R,0) \rangle_{\mathrm{eq}}$, rather than only the autocorrelation of the spatially averaged current. For homogeneous crystals, translational invariance reduces the kernel to a function of $\vec R-\vec r$, enabling efficient reciprocal-space representations. For heterogeneous or finite systems, coarse graining, symmetry reduction, and low-rank or basis-function representations provide practical routes to tractability.

\subsubsection{Spatially Resolved Local Conductive Energy-Flux Operator}
\label{sec:Hardy_flux}

To evaluate the space-projected correlations entering Eq.~\eqref{eq:gGK_full}, one requires a microscopic energy-flux field resolved in real space. A convenient construction is provided by the local energy-flux operator introduced by Hardy~\cite{hardy1963energy}, which yields a conserved energy current consistent with the microscopic continuity equation. As shown in Appendix~\ref{app:linear_response_hardy_J}, Hardy’s formulation leads naturally to a bond-centered representation of the conductive heat flux,
\begin{equation}
\label{eq:j_bond}
\vec j(\vec r) = -\sum_{ij} \vec R_{ij}\,\Psi_{ij}\, w(\vec r;\vec R_i,\vec R_j),
\end{equation}
with $\Psi_{ij} = \frac{1}{2}\left(\Xi_{ij}+\Xi_{ij}^\dagger\right)$ and $ \Xi_{ij} = \frac{1}{i\hbar}\left[\frac{p_i^2}{2m_i},V_j\right]$, and $w(\vec r;\vec R_i,\vec R_j) = \int_0^1 d\xi\, \Delta\!\left(\vec r-\big[(1-\xi)\vec R_i+\xi\vec R_j\big]\right)$ (spatial projection function). Because the localization function is normalized, spatial integration recovers the conventional bulk current. Equation~\eqref{eq:j_bond} therefore provides a direct route from atomistic trajectories to the two-point kernel in Eq.~\eqref{eq:gGK_full}, without invoking phonon modes, group velocities, or scattering models, and is consequently well suited to nonuniform media.

An instructive limiting case is a disordered harmonic lattice. In the appropriate diffusive or coarse-grained limit, the kernel constructed from Eq.~\eqref{eq:j_bond} reduces to a real-space representation of Allen--Feldman-type transport~\cite{Allen_Feldman_PhysRevB.48.12581}, in which energy transfer is governed by products of local heat-current matrix elements between nearly degenerate vibrational modes. The conventional Allen--Feldman conductivity is recovered only after spatial integration, whereas the present formulation retains the spatial structure of the underlying mode-coupling process. A detailed derivation is provided in Appendix~\ref{app:AF_kernel}.

In practice, the kernel is computationally realizable through controlled approximations at different levels of description. At the atomistic level, first-principles phonon properties provide mode-resolved lifetimes and velocities that define the kernel within the relaxation-time approximation as a superposition of exponentially attenuated streaming processes (Sec.~\ref{sec:numerical_simulation}). At larger scales, experimentally accessible observables, such as TTG Fourier components or time-domain thermoreflectance temporal response functions, constrain projected forms of the same kernel. The kernel framework therefore provides a unifying structure linking atomistic simulations and continuum descriptions, without introducing phenomenological boundary conditions.

\section{Numerical Illustration: Kernel-Based Transport in Bulk Silicon at 300~K}
\label{sec:numerical_simulation}

To illustrate the physical content of the kernel and validate internal consistency, we construct the spatiotemporal kernel for bulk crystalline silicon at 300~K within the RTA and apply it to TTG configurations spanning diffusive and quasi-ballistic regimes. As noted in Sec.~\ref{sec:unified_framework}, the RTA provides a complete realization of the formalism in which all kernel limits are available in closed form [Eqs.~\eqref{eq:rta_kernel_appendix} and \eqref{eq:RTA_kernel_qw}], enabling efficient continuum simulations without invoking frequency-domain Green's functions~\cite{hua2014analytical} or Monte Carlo solutions of the phBTE~\cite{peraud2011efficient}. We first analyze the kernel structure and then compute TTG responses across diffusive and quasi-ballistic regimes.

\subsection{Spatiotemporal Kernels in the RTA Limit}
\label{sec:numerical_kernel}

\begin{figure*}[!htbp]
\centering
\includegraphics[width=0.45\textwidth]{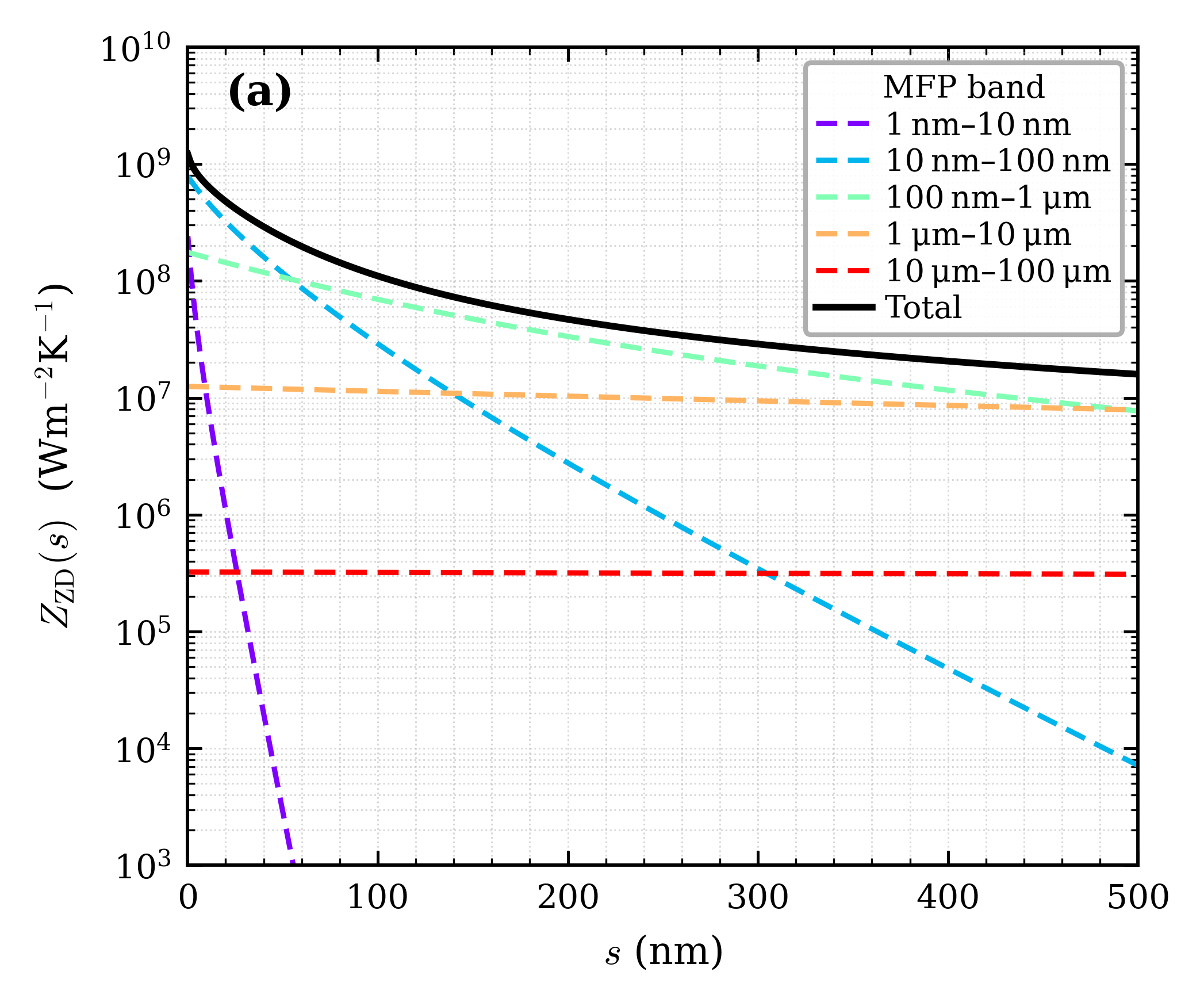}\hfill
\includegraphics[width=0.45\textwidth]{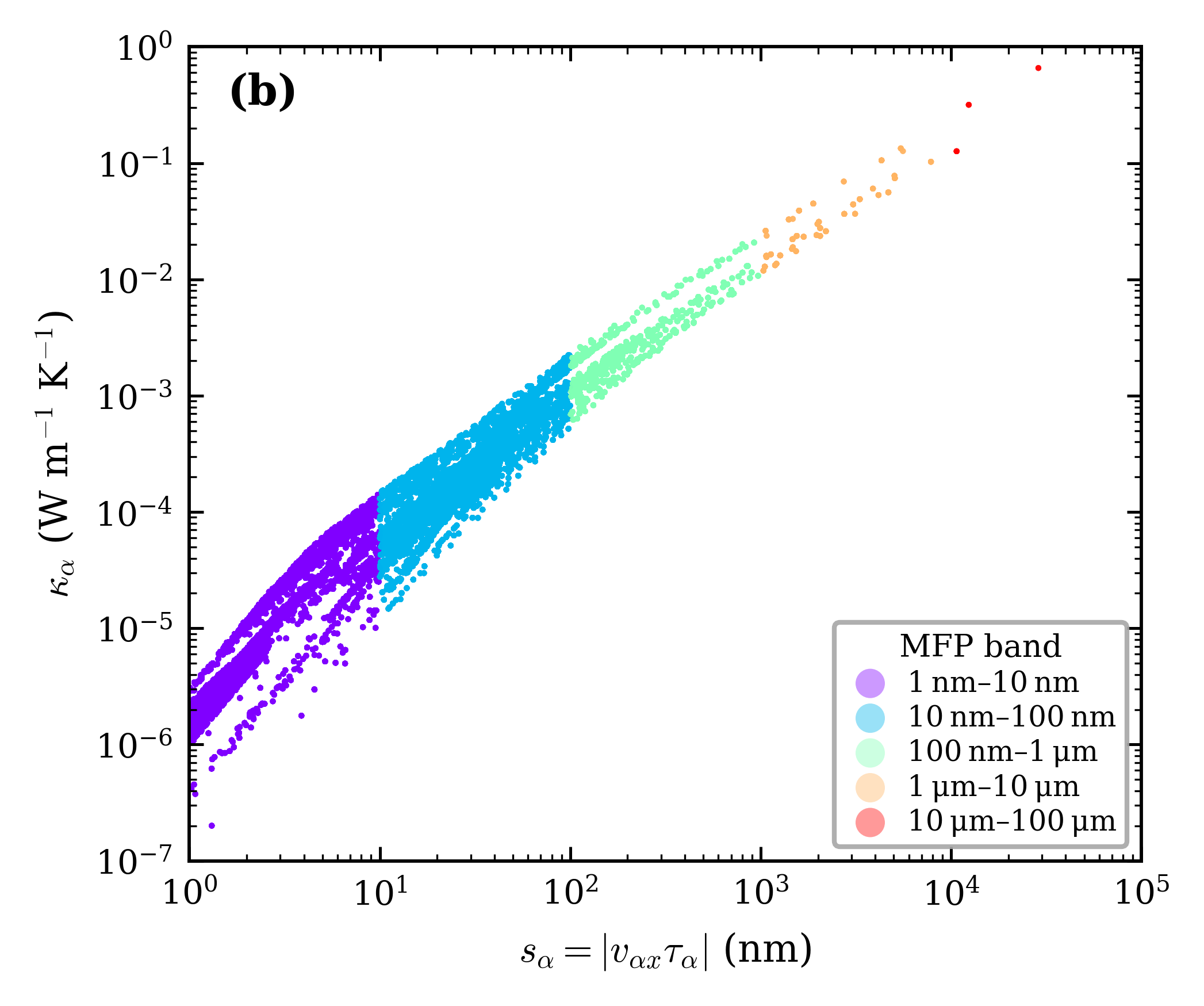}
\caption{Spatiotemporal transport characteristics of bulk silicon at 300~K. (a)~Zwanzig-diffusion kernel $Z_{\mathrm{ZD},xx}(x)$ along $[100]$, decomposed into contributions from phonon modes grouped by projected MFP (color-coded from short to long MFP in a cold-to-warm palette). The finite spatial extent reflects nonlocal transport; $\int dx\,Z_{\mathrm{ZD},xx}(x)=\kappa_{\mathrm{bulk}}$. (b)~Mode-resolved thermal conductivity $\kappa_\alpha$ versus projected MFP $\ell_{\alpha x}=|v_{\alpha x}\tau_\alpha|$, with each point representing a single phonon mode using the same MFP color coding as panel~(a).}
\label{fig:kernel_Si}
\end{figure*}

We begin with the RTA kernel constructed from first-principles phonon properties on a $41\times41\times41$ Brillouin-zone mesh (details in Ref.~\cite{crawford2024theory}). Because $\tau_\alpha$ controls both the temporal decay [Eq.~\eqref{eq:rta_LT}] and the spatial extent via $\vec{\ell}_\alpha=\vec v_\alpha\tau_\alpha$ [Eq.~\eqref{eq:rta_ZD}], memory and nonlocality are intrinsically coupled at the modal level. The Zwanzig-diffusion kernel $Z_{\mathrm{ZD},xx}(x)$ along $[100]$ extends over a finite range in $x$, set by the underlying MFP distribution, which spans several decades [Fig.~\ref{fig:kernel_Si}(a)]; decomposition by projected MFP $\ell_{\alpha x}=|v_{\alpha x}\tau_\alpha|$ reveals that different modes dominate different spatial scales, and the mode-resolved conductivity $\kappa_\alpha$ [Fig.~\ref{fig:kernel_Si}(b)] confirms this multiscale character, with long-MFP modes contributing disproportionately to transport. No single relaxation time or nonlocal length scale adequately characterizes the response. Instead, transport is governed by a broad spectrum of phonon MFP and relaxation times. When the characteristic length scale becomes comparable to the MFP of long-lived modes, their directional streaming leads to deviations from local constitutive behavior. In this regime, reduced descriptions based on single-parameter approximations become insufficient, and an accurate description requires resolving the underlying distribution encoded in the kernel.

\subsection{Transient Thermal Grating (TTG) Response}
\label{sec:numerical_TTG}

\begin{figure*}[!htbp]
\centering
\includegraphics[width=\textwidth]{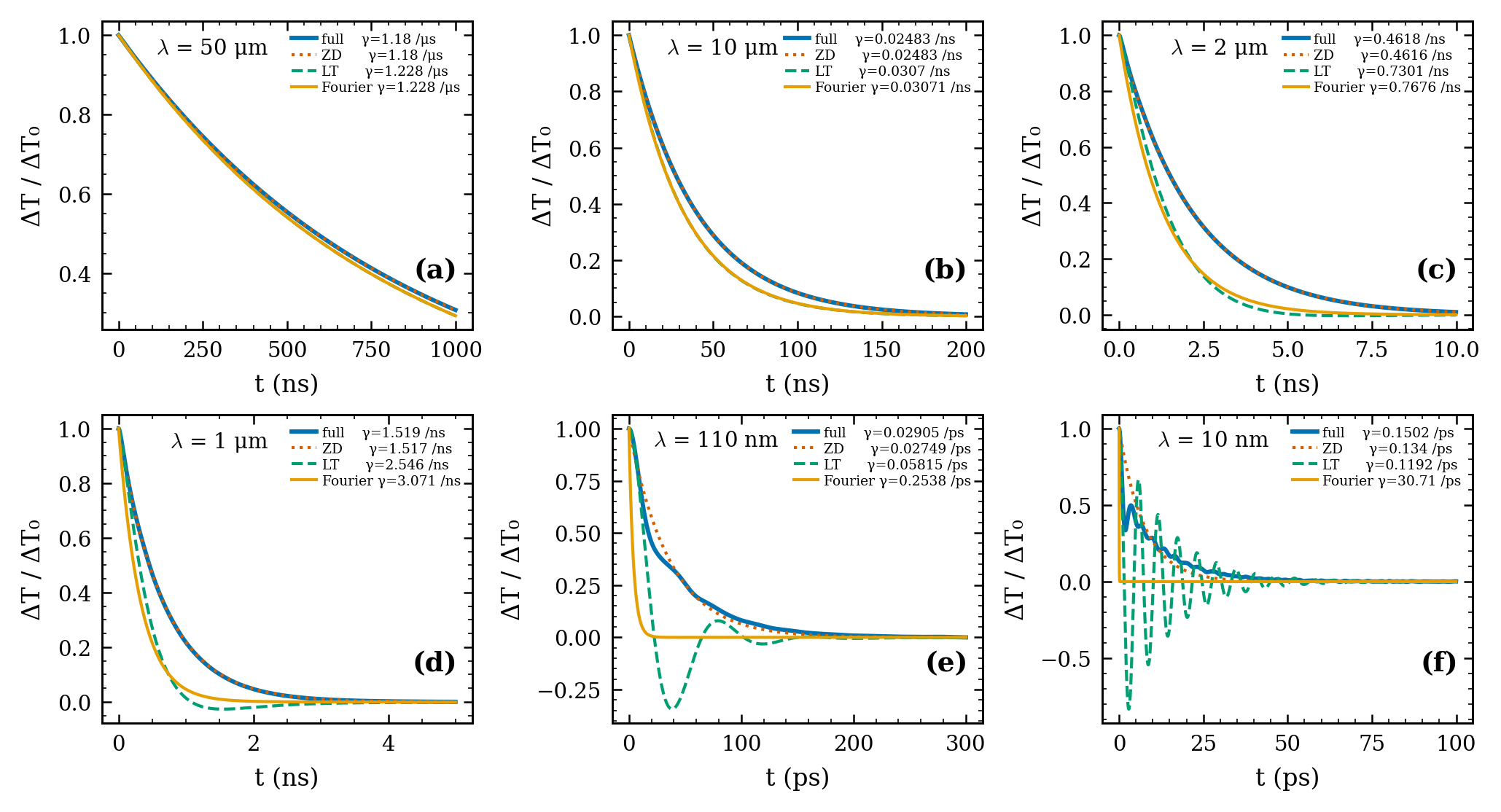}
\caption{Time-domain decay of the TTG amplitude in bulk silicon at 300~K for six grating periods $L$, spanning from diffusive (upper left) to deep quasi-ballistic (lower right) regimes. Each panel compares the full spatiotemporal kernel (solid), ZD (dotted), LT (dashed), and Fourier (solid) limits.}
\label{fig:ttg_decay}
\end{figure*}

In TTG experiments, a sinusoidal temperature profile with wavevector $q=2\pi/L$ is generated~\cite{johnson2013direct,maznev2011onset}. Because the perturbation is single-mode in space, the response depends on $\widetilde{\mathbb Z}(\vec q,\omega)$ [Eq.~\eqref{eq:RTA_kernel_qw}], providing a direct probe of scale-dependent transport. The grating amplitude decays from diffusive to deep quasi-ballistic behavior as $L$ decreases from $50~\mu$m to $10$~nm (Fig.~\ref{fig:ttg_decay}). At large $L$, all models collapse to Fourier behavior; as $L$ decreases toward the dominant MFP range, the full kernel and ZD predictions deviate from Fourier theory due to spatial nonlocality, while the LT model overestimates decay rates by neglecting spatial structure.

In the submicron regime, the hierarchy becomes clear. The LT model predicts pronounced oscillatory, second-sound-like behavior arising from temporal memory alone. However, these oscillations are strongly suppressed in the full kernel once spatial nonlocality is included: at $L=100$~nm, the peak-to-trough oscillation amplitude in the LT model exceeds 50\% of the initial signal, whereas the full kernel suppresses this to below 5\%, a tenfold reduction attributable to spatial dephasing among modes with different MFP. The ZD model, which retains spatial nonlocality, accurately captures the decay envelope except at very short times. Thus, spatial nonlocality governs the dominant departure from Fourier transport, while temporal memory primarily affects early-time dynamics. We note that a G--K-type (PH) description, which retains a single relaxation time and a single nonlocal length, would fall between the Fourier and ZD predictions in this hierarchy. For Si at 300~K, the broad MFP distribution (Fig.~\ref{fig:kernel_Si}(b)) precludes collapsing the kernel onto a single ($\tau$, $\ell$) pair; the full kernel or its ZD reduction is therefore required for quantitative accuracy across length scales, while a G--K fit would necessarily trade fidelity at one scale for accuracy at another.

This distinction has broader implications: second-sound-like oscillations in $T(\vec r,t)$ do not uniquely imply phonon hydrodynamics, but rather reflect the interplay between temporal memory and spatial coherence. In the LT limit, all modes share the same wavevector and interfere constructively because their spatial structure is projected out. Once spatial nonlocality is restored, each mode propagates over its own characteristic mean free path, so modes with widely different $\ell_\alpha$ accumulate different spatial phases and their contributions average destructively, giving rise to the tenfold suppression observed above. Hydrodynamic behavior corresponds to the special case in which a small number of collective modes dominate with well-separated time scales, allowing coherent wave propagation to persist.

These results are consistent with TTG experiments. Bencivenga \emph{et al.}~\cite{2019_EUV_TTG} observed clear deviations from Fourier diffusion in silicon at 300~K without oscillatory signatures, in agreement with the ZD prediction. The measured decay time ($\sim36$~ps) matches the kernel prediction, indicating that spatial nonlocality dominates under these conditions. Hydrodynamic-based analyses that employ an effective nonlocal length~\cite{beardo2022hydrodynamic,xiang2022time} can reproduce these observations at the level of reduced models; within the present framework, such a single length parameter corresponds to a coarse-grained projection of the underlying kernel, which is inherently multiscale. Specifically, the ZD kernel (Fig.~\ref{fig:kernel_Si} (a)) has significant weight extending to $\sim\!1~\mu$m, with a characteristic spatial extent of $100$--$200$~nm, consistent with the effective nonlocal lengths ($\ell\approx 100$--$200$~nm) fitted independently from TTG and TDTR experiments~\cite{beardo2022hydrodynamic,xiang2022time}, reinforcing the interpretation of G--K nonlocal lengths as reduced representations of the full kernel structure.

\begin{figure}[!htbp]
\centering
\includegraphics[width=3.375in]{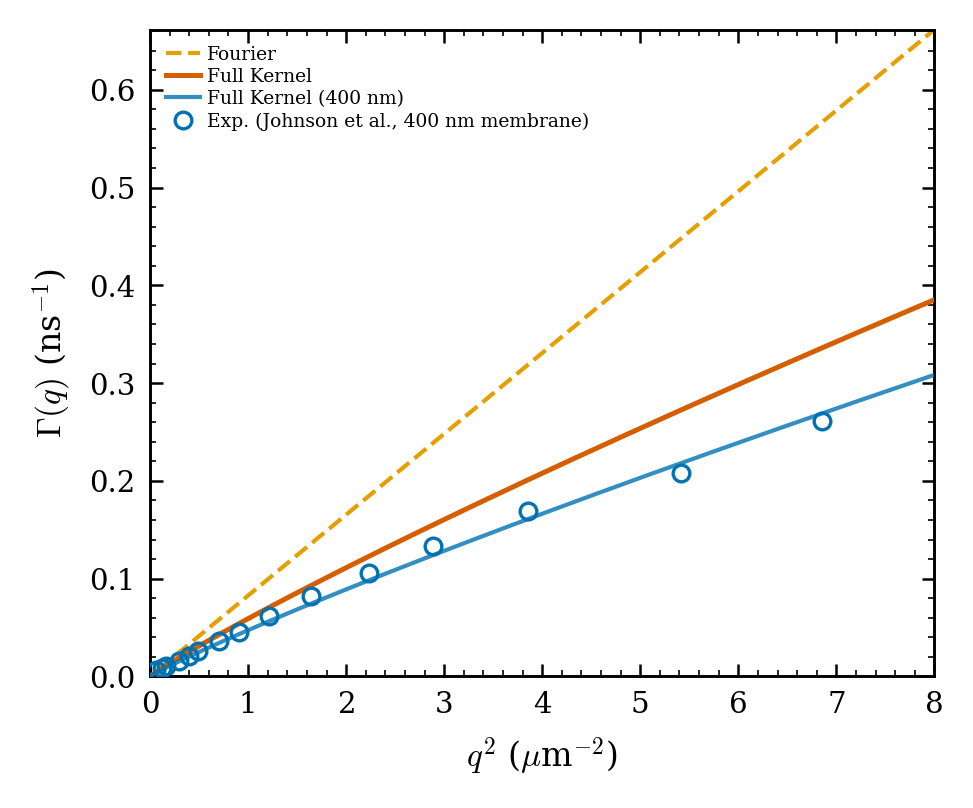}
\caption{TTG decay rate $\Gamma(q)$ versus $q^2$ for bulk silicon at 300~K (RTA). The dashed line is the Fourier limit $\Gamma=\alpha_{\mathrm{bulk}}q^2$, where $\alpha_{\mathrm{bulk}}=\kappa_{\mathrm{bulk}}/(\rho c)$ is the bulk thermal diffusivity. The dark solid curve is the full spatiotemporal kernel for bulk silicon, and the light solid curve includes the effect of boundary scattering in a 400~nm membrane. Open symbols denote experimental data from Johnson \emph{et al.}~\cite{johnson2013direct} for a free-standing 400~nm Si membrane.}
\label{fig:ttg_gamma_q2}
\end{figure}

The decay rate $\Gamma(q)$ deviates sublinearly from the Fourier relation $\Gamma=\alpha_{\mathrm{bulk}}q^2$ (where $\alpha_{\mathrm{bulk}}=\kappa_{\mathrm{bulk}}/(\rho c)$ is the bulk thermal diffusivity) at large $q$ (Fig.~\ref{fig:ttg_gamma_q2}), reflecting the suppression of long-MFP modes at short length scales. Including boundary scattering yields quantitative agreement with experiment, demonstrating how finite-size effects enter through selective truncation of long-range correlations. This boundary-scattering correction is introduced as an external empirical input rather than derived from the kernel itself; extending the formalism to incorporate finite-size boundary effects self-consistently is a natural direction for future work. The choice between mode-resolved and collective transport descriptions is geometry dependent. In one-dimensional TTG configurations, the sinusoidal perturbation resolves individual MFP contributions without angular averaging, and the mode-resolved RTA kernel captures this structure quantitatively. Beardo \emph{et al.}~\cite{beardo2022hydrodynamic} have shown that for two- and three-dimensional heater geometries on Si at 300~K, angular and spectral averaging effectively collapses the multiscale MFP spectrum onto a collective response characterized by a single nonlocal length ($\ell\approx190$~nm). Thus, the relative importance of mode-resolved versus collective descriptions depends on the measurement geometry, and the kernel formalism accommodates both through the choice of diagonal or off-diagonal mode-coupling structure. In summary, the spatiotemporal kernel naturally captures spatial nonlocality and temporal memory, while boundary scattering can be incorporated phenomenologically in the present implementation.

\section{Conclusion and Outlook}
\label{sec:conclusion}

We have formulated heat conduction in terms of a causal two-point spatiotemporal kernel $\overleftrightarrow{\mathbb{Z}}(\vec r,\vec R,t)$ derived from the Zwanzig projection-operator formalism and defined microscopically through a spatiotemporal Green--Kubo relation. Within this framework, Fourier diffusion, the Guyer--Krumhansl equation, and quasi-ballistic transport emerge as controlled asymptotic limits of a single constitutive object, providing a unified description across transport regimes without invoking separate phenomenological models. This formulation identifies temporal memory and spatial nonlocality as complementary manifestations of the same underlying microscopic dynamics, and provides a systematic basis for organizing transport behavior beyond local equilibrium descriptions.

The analysis clarifies the origin of wave-like thermal response. In particular, oscillatory features in $T(\vec r,t)$ arise from the interplay of temporal memory and spatial coherence, rather than from hydrodynamic collective behavior alone. For silicon at 300~K, the broad distribution of phonon MFP leads to strong spatial dephasing that suppresses coherent second-sound-like oscillations, indicating that hydrodynamic transport represents a special limiting case rather than a generic consequence of nonlocality. Because the kernel is defined through equilibrium heat-flux correlations, the formalism is not restricted to the phonon-gas picture; for disordered harmonic solids, it recovers a spatial diffusion kernel consistent with the Allen--Feldman limit~\cite{Allen_Feldman_PhysRevB.48.12581}.

To illustrate the framework, we constructed the kernel for crystalline silicon at room temperature within the relaxation-time approximation and applied it to transient thermal grating configurations. Under these conditions, spatial nonlocality associated with the phonon mean-free-path distribution provides the dominant contribution to deviations from Fourier transport, while temporal memory primarily influences short-time dynamics. The Zwanzig-diffusion reduction captures the observed decay behavior and $\Gamma(q)$ dispersion over a broad range of length scales~\cite{2019_EUV_TTG,johnson2013direct}, whereas purely local transient models do not. These results demonstrate how different transport regimes can be interpreted within a single constitutive framework, with reduced descriptions arising in a controlled manner from the underlying kernel.

Several challenges remain for practical implementation. The present numerical illustration is restricted to crystalline Si within the RTA, the simplest realization of the formalism. Among the extensions developed here, the spatiotemporal Green--Kubo evaluation via equilibrium molecular dynamics (for disordered or composite systems) and the construction of first-principles interfacial kernels (for heterogeneous media) are the most immediate targets for quantitative validation; both are the subject of ongoing work. Efficient representations of the kernel, such as low-rank or separable approximations, will be important for large-scale simulations. The derivation of consistent boundary conditions for nonlocal constitutive relations remains an open problem; while such conditions are known for reduced models such as the G--K equation~\cite{guo2018phonon}, their extension to the full kernel framework is not yet established. The present formulation does not include effects associated with quantum phase coherence, such as phonon interference at superlattice interfaces.

From an experimental perspective, the kernel can be viewed as a constitutive response function whose projections are accessed across different measurement platforms. Conventional transport coefficients correspond to strongly coarse-grained limits, while ultrafast and nanoscale measurements probe its temporal and spatial structure more directly. Because the kernel is defined as a time-correlation function of equilibrium heat-flux fluctuations, it satisfies the positivity conditions required for non-negative entropy production, providing a microscopic basis for thermodynamic consistency beyond local Fourier transport.

More broadly, this work supports the view that spatiotemporal response functions, rather than transport coefficients alone, provide a natural constitutive description of nonequilibrium thermal transport. In this perspective, classical diffusion emerges as a coarse-grained limit of a more general causal response that retains memory, nonlocality, and heterogeneity. The present formulation establishes a framework for connecting microscopic dynamics to such generalized constitutive behavior in a systematic and physically transparent manner.

\begin{acknowledgments}
This work was authored in part by the National Laboratory of the Rockies for the U.S. Department of Energy (DOE), operated under Contract No. DE-AC36-08GO28308. Funding provided by the U.S. Department of Energy Office of Science Energy Earthshots Initiative as part of the Degradation Reactions in Electrothermal Energy Storage (DEGREES) project. The views expressed in the article do not necessarily represent the views of the DOE or the U.S. Government. The U.S. Government retains and the publisher, by accepting the article for publication, acknowledges that the U.S. Government retains a nonexclusive, paid-up, irrevocable, worldwide license to publish or reproduce the published form of this work, or allow others to do so, for U.S. Government purposes. J.D. and Y.Z. also acknowledge discussions with Dr. J. Vidal and Ben Tzou.
\end{acknowledgments}

\appendix

%%%%%%%%%%%%%%%%%%%%%%%%%%%%%%%%%%%%%%%%%%%%%%%%%%%%%%%%%%
\section{Kinetic Phonon Transport Theory}
\label{app:model-coupled-dynamics}
%%%%%%%%%%%%%%%%%%%%%%%%%%%%%%%%%%%%%%%%%%%%%%%%%%%%%%%%%%

\subsection{Local Thermal Equilibrium in a Bulk Phonon Gas}
\label{app:phonon}
In thermal equilibrium at temperature $T_0$, the average occupation number of phonon mode $\alpha$ is given by the Bose--Einstein distribution,$n_\alpha^{\mathrm{eq}}(T_0) = \frac{1}{e^{\hbar\omega_\alpha/k_\mathrm{B}T_0}-1}$, with $\hbar$ being the reduced Planck constant and $k_\mathrm{B}$ being the Boltzmann constant. The corresponding equilibrium energy density is $e^{\mathrm{eq}}(T_0) = \frac{1}{V} \sum_{\alpha} \hbar\omega_\alpha \,n_\alpha^{\mathrm{eq}}(T_0)$. Thermal fluctuations cause the instantaneous phonon occupations to deviate from their equilibrium values. For a bosonic mode $\alpha$, the equilibrium variance is $ \sigma_\alpha^2(T_0) = n_\alpha^{\mathrm{eq}}(T_0) \left[ n_\alpha^{\mathrm{eq}}(T_0)+1 \right] = \frac{1}{4} \left[ \sinh\!\left( \frac{\hbar\omega_\alpha}{2k_\mathrm{B}T_0} \right) \right]^{-2}$, and these fluctuations are directly related to the mode-resolved heat capacity:
\begin{equation}
\label{eq:mode_c}
c_\alpha(T_0) = \hbar\omega_\alpha \left. \frac{d n_\alpha^{\mathrm{eq}}(T)}{dT} \right|_{T_0} = k_\mathrm{B} \left( \frac{\hbar\omega_\alpha/2k_\mathrm{B}T_0} {\sinh(\hbar\omega_\alpha/2k_\mathrm{B}T_0)} \right)^2 = \frac{(\hbar\omega_\alpha)^2}{k_\mathrm{B}T_0^2} \sigma_\alpha^2(T_0).
\end{equation}
The volumetric heat capacity is therefore $\rho c = \frac{C}{V}=\frac{1}{V} \sum_{\alpha} c_\alpha(T_0)$.
%\rho c = \left. \frac{d e^{\mathrm{eq}}(T)}{dT} \right|_{T_0} = \frac{1}{V} \sum_{\alpha} c_\alpha(T_0).
For small deviations from equilibrium, the local temperature field $T(\vec r,t)$ may be defined by linearizing the thermodynamic relation between energy density and temperature about the reference state [Eq. \eqref{eq:OP_local_T}], showing that the local temperature deviation is determined by the energy-weighted deviation of the phonon occupations, normalized by the total heat capacity.
%\begin{equation}
%\label{eq:T_local_var}
%T(\vec r,t) = T_0 + \frac{1}{\rho c} \left[ e(\vec r,t)-e^{\mathrm{eq}}%(T_0) \right].
%\end{equation}
%where $\rho c$ is the volumetric heat capacity.
%Substituting Eqs.~\eqref{eq:e_eq} and \eqref{eq:rhoC} into Eq.~\eqref{eq:T_local_var} yields the microscopic expression
%\begin{equation}
%\label{eq:T_local_var2}
%T(\vec r,t) = T_0 + \frac{ \sum_{\alpha} \hbar\omega_\alpha \left[ n_\alpha(\vec r,t)-n_\alpha^{\mathrm{eq}}(T_0) \right] }{ \sum_{\alpha} c_\alpha(T_0) }.
%\end{equation}
%Equation~\eqref{eq:T_local_var2} shows t

\subsection{Causal Integral Solutions and Inhomogeneous Generalizations}
\label{sec:causal_solutions}

This subsection develops the causal integral solutions and inhomogeneous generalizations of the Zwanzig-projected phBTE whose coupled differential form is given in Eqs.~\eqref{eq:bte_r}. We consider both homogeneous phonon gases, in which all microscopic parameters are spatially uniform, and inhomogeneous media, in which the local phonon spectrum and scattering operator vary with position.

%%%%%%%%%%%%%%%%%%%%%%%%%%%%%%%%%%%%%%%%%%%%%%%%%%%%%%%%%%
\subsubsection{Homogeneous Phonon Gases}
\label{app:homogeneous_phonon}
%%%%%%%%%%%%%%%%%%%%%%%%%%%%%%%%%%%%%%%%%%%%%%%%%%%%%%%%%%

To make the causal, memory-dependent structure of the homogeneous coupled equations [Eqs.~\eqref{eq:bte_r}] explicit, we remove the homogeneous exponential decay by defining $ \tilde{\eta}_{\lambda} (\vec r , t) = e^{\gamma_{\lambda} t} \eta_{\lambda} (\vec r, t)$ and $ \tilde{\theta}_{\iota} (\vec r , t) = e^{\varepsilon_{\iota} t} \theta_{\iota} (\vec r, t)$. With these definitions, Eqs.~\eqref{eq:bte_O_r}--\eqref{eq:bte_E_r} become
\begin{subequations}
\label{eq:bte_tilde_revised}
\begin{eqnarray}
\label{eq:bte_eta_tilde_revised}
\frac{\partial \tilde{\eta}_{\lambda}(\vec r , t)}{\partial t} &=& -\frac{e^{\gamma_{\lambda} t}}{k_\text{B} T^{2}_0} \vec{\Lambda}_\lambda \cdot \nabla T(\vec r, t) - e^{\gamma_{\lambda} t} \sum_{\iota=1}^{{M}_{\text{E}}} e^{-\varepsilon_{\iota} t}\vec{{\Pi}}_{\lambda\iota} \cdot \nabla \tilde{\theta}_{\iota} (\vec r,t), \nonumber \\
\label{eq:bte_theta_tilde_revised}
\frac{\partial \tilde{\theta}_{\iota}(\vec r , t)}{\partial t} &=& - e^{\varepsilon_{\iota} t} \sum_{\lambda =1}^{{M}_{\text{O}}} e^{-\gamma_{\lambda} t}\vec{{\Pi}}_{\lambda\iota} \cdot \nabla \tilde{\eta}_{\lambda} (\vec r,t).
\end{eqnarray}
\end{subequations}
Assuming stability as $t\to -\infty$, these equations admit causal convolution solutions. Restoring the original variables gives
\begin{subequations}
\label{eq:ana2_revised}
\begin{align}
\eta_\lambda(\vec r, t) 
&= - \int_{0}^{\infty} dt_1 \frac{e^{-\gamma_{\lambda}t_1}}{k_\text{B} T_0^2} \, \vec{\Lambda}_\lambda \cdot \nabla T(\vec r, t-t_1) \nonumber \\
&\quad - \int_{0}^{\infty} dt_1 \, e^{-\gamma_{\lambda}t_1} \sum_{\iota=1}^{M_{\text{E}}} \vec{\Pi}_{\lambda\iota} \cdot \nabla \theta_\iota(\vec r, t-t_1),
\label{eq:ana_2a_revised}
\\
\theta_{\iota}(\vec r, t) 
&= - \int_{0}^{\infty} dt_1 \, e^{-\varepsilon_\iota t_1} \sum_{\lambda=1}^{M_{\text{O}}} \vec{\Pi}_{\lambda\iota} \cdot \nabla \eta_{\lambda}(\vec r, t-t_1).
\label{eq:ana_2b_revised}
\end{align}
\end{subequations}

Equations~\eqref{eq:ana2_revised} summarize the central structure of the Zwanzig-projected phBTE: the temperature is the sole independent macroscopic transport variable, while all auxiliary fields respond through time-retarded couplings. Spatial nonlocality enters through the recursive gradient couplings between $\eta_\lambda$ and $\theta_\iota$. Eliminating $\theta_\iota$ by substituting Eq.~\eqref{eq:ana_2b_revised} into Eq.~\eqref{eq:ana_2a_revised} yields a closed equation for the flux-mode amplitudes,
\begin{equation}
\label{eq:flux-mode-only_revised}
\begin{split}
\eta_{\lambda}(\vec r, t) &= -\int_{0}^{\infty} dt_1 \frac{e^{-\gamma_\lambda t_1}}{k_\text{B} T^{2}_0} \, \vec{\Lambda}_\lambda \cdot \nabla T(\vec r , t-t_1) \\
&\quad + \int_{0}^{\infty} dt_1\, e^{-\gamma_\lambda t_1} \sum_{\lambda'=1}^{M_{\text{O}}} \int_{0}^{\infty} dt_2 \sum_{I,J=1}^{3} \phi_{\lambda \lambda' I J}(t_2)\, \partial^2_{IJ}\, \eta_{\lambda'} (\vec r, t-t_1-t_2),
\end{split}
\end{equation}
where
\begin{equation}
\label{eq:phi_revised}
\phi_{\lambda \lambda' I J} (t) = \sum_{\iota=1}^{M_{\text{E}}} \Pi_{\lambda\iota,I}\, \Pi_{\lambda'\iota,J}\, e^{-\varepsilon_\iota t}.
\end{equation}
Equation~\eqref{eq:flux-mode-only_revised} shows that the even-parity nonlocal modes can be formally eliminated, but their influence persists through a non-Markovian closure: the evolution of $\eta_\lambda$ depends not only on $\nabla T$ but also on second spatial derivatives of $\eta_{\lambda'}$.

%%%%%%%%%%%%%%%%%%%%%%%%%%%%%%%%%%%%%%%%%%%%%%%%%%%%%%%%%%
\subsubsection{Inhomogeneous Phonon Gases}
\label{sec:inhomogeneous_phonon}
%%%%%%%%%%%%%%%%%%%%%%%%%%%%%%%%%%%%%%%%%%%%%%%%%%%%%%%%%%

Real materials may exhibit spatially varying composition, strain, atomic structure, or finite-size confinement. When such variation is smooth on the scale of the lattice spacing, the phonon gas description can be generalized by allowing the local phonon spectrum and scattering operator to depend on position, $\mathcal{L}\rightarrow\mathcal{L}(\vec r)$. Within the Zwanzig framework, this corresponds to a local eigenmode decomposition at each $\vec r$, leading to position-dependent relaxation rates and coupling coefficients. The causal structure of the homogeneous theory is preserved, but additional terms arise from gradients of the microscopic coefficients themselves. The inhomogeneous generalization of Eqs.~\eqref{eq:ana2_revised} takes the form
\begin{subequations}
\label{eq:eta_theta_r_revised}
\begin{eqnarray}
\label{eq:eta_r_revised}
\eta_\lambda(\vec r, t) = &-& \int_{0}^{\infty} dt_1 \frac{e^{-\gamma_{\lambda}(\vec r)t_1}}{k_\text{B} T_0^2}
\, \vec{\Lambda}_\lambda(\vec r) \cdot \nabla T(\vec r, t-t_1)  \nonumber\\
&-& \int_{0}^{\infty} dt_1 \, e^{-\gamma_{\lambda}(\vec r)t_1} \sum_{\iota=1}^{M_{\text{E}}} \vec{\Pi}_{\lambda\iota}(\vec r) \cdot \nabla \theta_\iota(\vec r, t-t_1), \\
\label{eq:theta_r_revised}
\theta_{\iota}(\vec r, t)= &-& \int_{0}^{\infty} dt_1 \, e^{-\varepsilon_\iota(\vec r) t_1} \sum_{\lambda=1}^{M_{\text{O}}} \vec{\Pi}_{\lambda\iota}(\vec r) \cdot \nabla \eta_{\lambda}(\vec r, t-t_1).
\end{eqnarray}
\end{subequations}
Applying $\nabla$ to Eq.~\eqref{eq:theta_r_revised} generates two contributions: (i) the intrinsic nonlocal term involving gradients of $\eta_\lambda$, and (ii) an inhomogeneity-induced term arising from gradients of $\vec{\Pi}_{\lambda\iota}(\vec r)$ and $\varepsilon_\iota(\vec r)$. Substituting back into Eq.~\eqref{eq:eta_r_revised} yields
\begin{equation}
\label{eq:flux-mode-only_r_revised}
\begin{split}
\eta_{\lambda}(\vec r, t) =& -\int_{0}^{\infty} dt_1 \frac{e^{-\gamma_\lambda(\vec r) t_1}}{k_\text{B} T^{2}_0} \, \vec{\Lambda}_\lambda(\vec r) \cdot \nabla T(\vec r , t-t_1) \\
&+ \int_{0}^{\infty} dt_1\, e^{-\gamma_\lambda(\vec r) t_1} \sum_{\lambda'=1}^{M_{\text{O}}} \int_{0}^{\infty} dt_2 \biggr[ \sum_{I,J=1}^{3} \Phi_{\lambda \lambda' I J}(\vec r,t_2)\, \partial^2_{IJ}\, \eta_{\lambda'} (\vec r, t-t_1-t_2)\\
&+ \sum_{I=1}^{3} \Psi_{\lambda \lambda' I}(\vec r,t_2)\, \partial_{I}\, \eta_{\lambda'} (\vec r, t-t_1-t_2) \biggr],
\end{split}
\end{equation}
where
\begin{equation}
\label{eq:Phi_r_revised}
\Phi_{\lambda \lambda' I J}(\vec r,t) = \sum_{\iota=1}^{M_{\text{E}}} \Pi_{\lambda\iota,I}(\vec r)\, \Pi_{\lambda'\iota,J}(\vec r)\, e^{-\varepsilon_\iota(\vec r) t},
\end{equation}
and
\begin{equation}
\label{eq:Psi_r_revised}
\Psi_{\lambda \lambda' I}(\vec r,t) = \sum_{\iota=1}^{M_{\text{E}}} \sum_{J=1}^{3} \Pi_{\lambda\iota,J}(\vec r)\, \frac{\partial}{\partial x_J} \biggr( \Pi_{\lambda'\iota,I}(\vec r)\,e^{-\varepsilon_{\iota}(\vec r) t} \biggr).
\end{equation}
Note that the uppercase $\Phi_{\lambda\lambda'IJ}(\vec r,t)$ and $\Psi_{\lambda\lambda'I}(\vec r,t)$ generalize the position-independent coefficients $\phi_{\lambda\lambda'IJ}(t)$ [Eq.~\eqref{eq:phi_revised}] to spatially varying media.
The coefficients $\Phi$ encode intrinsic nonlocality that persists even in homogeneous media and produce second-derivative spatial couplings. By contrast, $\Psi$ arises entirely from spatial variation of the microscopic coefficients and vanishes in homogeneous systems. It therefore represents the leading correction associated with inhomogeneity.

Taken together, the homogeneous and inhomogeneous Zwanzig-projected equations provide a controlled route from microscopic phonon dynamics to reduced transport descriptions. In Appendix~\ref{app:full_kernel}, we reorganize this hierarchy into a compact kernel representation in space and time that interfaces directly with continuum modeling.

\subsubsection{Iterative Expansion of Modal Solutions}
\label{app:iterative_solution}
The iterative expansion of Eq.~\eqref{eq:flux-mode-only_r_revised} allows systematic inclusion of memory and spatial nonlocality. For example, the flux-mode variable $\eta_\lambda(\vec{r},t)$ can be written as: $\eta_\lambda =  \eta^{(0)}_\lambda + \Delta \eta^{(1)}_\lambda + \Delta \eta^{(2)}_\lambda + \dots~,$
where the zeroth-order term $\eta^{(0)}_\lambda$ is
\begin{equation}
\label{eq:eta0_r}
\eta^{(0)}_\lambda (\vec{r}, t) = - \frac{1}{k_\mathrm{B} T^2_0} \int_{0}^{\infty} dt_1 e^{-\gamma_{\lambda}(\vec{r})t_1}  \biggr( \vec{\Lambda}_\lambda(\vec{r}) \cdot \vec{\nabla}_{r}T(\vec{r}, t-t_1) \biggr),  
\end{equation}
and the first-order correction $\Delta \eta^{(1)}_\lambda$ is
\begin{equation}
\label{eq:eta1_r}
\begin{split}
\Delta \eta^{(1)}_\lambda (\vec{r}, t) =& -\frac{1}{k_\mathrm{B}T_0^2}\int_{0}^{\infty} dt_1  \int_{0}^{\infty} dt_2 \int_{0}^{\infty} dt_3 e^{-\gamma_\lambda(\vec{r}) t_1} \sum_{\lambda'=1}^{M_{\text{O}}}  \\
\biggr(&\sum_{I=1}^{3} \Psi_{\lambda\lambda'I}(\vec{r},t_2) \frac{\partial}{\partial x_I} [e^{-\gamma_{\lambda'}(\vec{r}) t_3} \vec{\Lambda}_{\lambda'}(\vec{r})  \cdot \vec{\nabla}_rT(\vec{r}, t-t_1-t_2-t_3)]\\
+ &\sum_{I,J=1}^{3} \phi_{\lambda\lambda'IJ}(t_2) \frac{\partial^2 }{\partial x_I \partial x_J} [e^{-\gamma_{\lambda'}(\vec{r}) t_3} \vec{\Lambda}_{\lambda'}(\vec{r})  \cdot \vec{\nabla}_rT(\vec{r}, t-t_1-t_2-t_3) ] \biggr).
\end{split}
\end{equation}
Next, we introduce the total lag $t'=t_1+t_2+t_3$ and keep $t_2$ and $t_3$ as inner time variables. The mapping $(t_1,t_2,t_3) \rightarrow (t',t_2,t_3)$ has unit Jacobian, i.e.\ $dt_1dt_2dt_3 = dt'dt_2dt_3$. Domain mapping of $t_1\ge0$, $t_2\ge0$, and $t_3\ge0$ gives:
\begin{equation}
\label{eq:domain}
t' \ge0,\qquad 0 \le t_2 \le t', \qquad 0 \le t_3 \le t' - t_2.  
\end{equation}
Hence, 
\begin{equation}
\label{eq:eta1_r_single}
\begin{split}
\Delta \eta^{(1)}_\lambda (\vec{r}, t) =& -\int_{0}^{\infty} dt' \frac{e^{-\gamma_\lambda(\vec{r}) t'}}{k_\mathrm{B}T_0^2} \int_{0}^{t'} dt_2 \int_{0}^{t'-t_2} dt_3 e^{\gamma_\lambda(\vec{r}) (t_2+t_3)} \sum_{\lambda'=1}^{M_{\text{O}}}  \\
&\biggr(\sum_{I=1}^{3} \Psi_{\lambda\lambda'I}(\vec{r},t_2) \frac{\partial}{\partial x_I} [ e^{-\gamma_{\lambda'}(\vec{r}) t_3} \vec{\Lambda}_{\lambda'}(\vec{r}) \cdot \vec{\nabla}_rT(\vec{r}, t-t') ]\\
&+  \sum_{I,J=1}^{3} \phi_{\lambda\lambda'IJ}(t_2) \frac{\partial^2 }{\partial x_I \partial x_J} [e^{-\gamma_{\lambda'}(\vec{r}) t_3} \vec{\Lambda}_{\lambda'}(\vec{r}) \cdot \vec{\nabla}_rT(\vec{r}, t-t') ] \biggr)\\
=& -\int_{0}^{\infty} dt' \frac{e^{-\gamma_\lambda (\vec{r})t'}}{k_\mathrm{B}T_0^2} \biggr( \vec{A}_0^{(1)}(\vec{r},t';\lambda) \cdot  \vec{\nabla}_rT(\vec{r}, t-t') +\sum_{I=1}^{3}  \vec{A}_1^{(1)}(\vec{r},t';\lambda,I) \cdot  \frac{\partial }{\partial x_I}\vec{\nabla}_rT(\vec{r}, t-t') \\
& + \sum_{I,J=1}^{3}  \vec{A}_2^{(1)}(\vec{r},t';\lambda,I,J) \cdot  \frac{\partial^2 }{\partial x_I \partial x_J}\vec{\nabla}_rT(\vec{r}, t-t') \biggr), 
\end{split}
\end{equation}
where
\begin{equation}
\label{eq:A0_1}
\begin{split}
\vec{A}_0^{(1)}(\vec{r},t';\lambda) =& \int_{0}^{t'} dt_2 \int_{0}^{t'-t_2} dt_3 e^{\gamma_\lambda(\vec{r}) (t_2+t_3)} \sum_{\lambda'=1}^{M_{\text{O}}}  \biggr( \\
& \sum_{I=1}^{3} \{\frac{\partial }{\partial x_I}[\Psi_{\lambda\lambda'I}(\vec{r},t_2)  e^{-\gamma_{\lambda'}(\vec{r}) t_3} \vec{\Lambda}_{\lambda'}(\vec{r})]-\frac{\partial \Psi_{\lambda\lambda'I}(\vec{r},t_2) }{\partial x_I}  e^{-\gamma_{\lambda'}(\vec{r}) t_3} \vec{\Lambda}_{\lambda'}(\vec{r})\} +\\
&\sum_{I,J=1}^{3} \{\frac{\partial^2 }{\partial x_I \partial x_J}[\phi_{\lambda\lambda'IJ}(\vec{r},t_2)  e^{-\gamma_{\lambda'}(\vec{r}) t_3} \vec{\Lambda}_{\lambda'}(\vec{r})]-\frac{\partial^2 \phi_{\lambda\lambda'IJ}(\vec{r},t_2)}{\partial x_I \partial x_J}  e^{-\gamma_{\lambda'}(\vec{r}) t_3} \vec{\Lambda}_{\lambda'}(\vec{r})\} \biggr),
\end{split}
\end{equation}

\begin{equation}
\label{eq:A1_1}
\vec{A}_1^{(1)}(\vec{r},t';\lambda,I) = \int_{0}^{t'} dt_2 \int_{0}^{t'-t_2} dt_3 e^{\gamma_\lambda(\vec{r}) (t_2+t_3)} \sum_{\lambda'=1}^{M_{\text{O}}} \Psi_{\lambda\lambda'I}(\vec{r},t_2)  e^{-\gamma_{\lambda'}(\vec{r}) t_3} \vec{\Lambda}_{\lambda'}(\vec{r}),
\end{equation}

\begin{equation}
\label{eq:A2_1}
\vec{A}_2^{(1)}(\vec{r},t';\lambda,I,J) = \int_{0}^{t'} dt_2 \int_{0}^{t'-t_2} dt_3 e^{\gamma_\lambda(\vec{r}) (t_2+t_3)} \sum_{\lambda'=1}^{M_{\text{O}}} \phi_{\lambda\lambda'IJ}(\vec{r},t_2)  e^{-\gamma_{\lambda'}(\vec{r}) t_3} \vec{\Lambda}_{\lambda'}(\vec{r}),
\end{equation}
Therefore, the first-order correction $\Delta \eta^{(1)}_\lambda (\vec{r}, t)$ depends on a time convolution of the temperature gradient field $\vec{\nabla}_r T(\vec{r}, t)$, its first-order spatial derivatives $\frac{\partial}{\partial x_I} \vec{\nabla}_r T(\vec{r}, t)$, and its second-order spatial derivatives $\frac{\partial^2}{\partial x_I \partial x_J} \vec{\nabla}_r T(\vec{r}, t)$. Although Eq.~\eqref{eq:eta1_r_single} originally involves three nested time integrals, this structure can be interpreted as a single causal convolution, whose kernel is given by the ordinary time-domain convolution of the corresponding nested integral kernels shown in Eqs.~\eqref{eq:A0_1}, \eqref{eq:A1_1}, and \eqref{eq:A2_1}.

The second-order correction $\Delta \eta^{(2)}_\lambda (\vec{r}, t)$ takes an analogous form,
\begin{equation}
\label{eq:eta2_r_single}
\begin{split}
\Delta \eta^{(2)}_\lambda (\vec{r}, t) =& -\int_{0}^{\infty} dt' \frac{e^{-\gamma_\lambda (\vec{r})t'}}{k_\mathrm{B}T_0^2} \biggr( \vec{A}_0^{(2)}(\vec{r},t';\lambda) \cdot  \vec{\nabla}_rT(\vec{r}, t-t')\\ &+\sum_{I=1}^{3}  \vec{A}_1^{(2)}(\vec{r},t';\lambda,I) \cdot  \frac{\partial }{\partial x_I}\vec{\nabla}_rT(\vec{r}, t-t') \\
& + \sum_{I,J=1}^{3}  \vec{A}_2^{(2)}(\vec{r},t';\lambda,I,J) \cdot  \frac{\partial^2 }{\partial x_I \partial x_J}\vec{\nabla}_rT(\vec{r}, t-t') \\
& + \sum_{I,J,K=1}^{3}  \vec{A}_3^{(2)}(\vec{r},t';\lambda,I,J,K) \cdot  \frac{\partial^3 }{\partial x_I \partial x_J \partial x_K}\vec{\nabla}_rT(\vec{r}, t-t') \biggr).
\end{split}
\end{equation}
Consequently, the macroscopic heat flux $\vec{j}(\vec{r}, t)$ is obtained by summing over the microscopic flux-modes. The iterative expansion for $\eta_\lambda$ leads to a generalized constitutive law:
\begin{equation}
\label{eq:J_expansion_r}
\vec{j}(\vec{r},t)  = - \sum_{n_1, n_2, n_3=0}^{\infty} \frac{1}{(n_1)! (n_2)! (n_3)!} \int_{0}^{\infty} dt' \overleftrightarrow{Z}^{(n_1+n_2+n_3)}(\vec{r}, t'; n_1, n_2, n_3)
\cdot  \frac{\partial^{(n_1+n_2+n_3)}} {\partial^{n_1} x \partial^{n_2} y \partial^{n_3} z} \vec{\nabla}_r T(\vec{r},t-t').
\end{equation}
Here $\overleftrightarrow{Z}^{(n)}(\vec{r}, t'; n_1, n_2, n_3)$ are single-time convolution kernels encapsulating higher-order memory and nonlocal effects.

This compact representation highlights the hierarchical nonlocal and memory-dependent structure of the transport response encoded in the correction term. The lowest-order term ($n_1=n_2=n_3=0$) yields the local transient heat flux,
\begin{equation}
\label{eq:total_local_j}
\vec{j}^{\text{local}}(\vec{r},t) = -\int_{0}^{\infty} dt' \, \overleftrightarrow{Z}^{(0)}(\vec{r},t';0,0,0) \cdot \vec{\nabla}_r T(\vec{r}, t-t').
\end{equation}
All higher-order terms ($n = n_1+n_2+n_3 \ge 1$) represent nonlocal corrections.

\subsection{Hydrodynamic Limit and Guyer--Krumhansl Equation}
\label{app:GK}

We first derive the hydrodynamic reduction of the Zwanzig-projected equations, then obtain the corresponding closed-form spatiotemporal kernel.

\subsubsection{Reduction}

In the phonon hydrodynamic regime~\cite{guyer1966solution,guyer1966thermal}, a separation of time scales allows the even-parity modes to be treated as fast variables slaved to the slow dynamics of the three odd-parity hydrodynamic modes ($\lambda=1,2,3$). The projected equations reduce to
\begin{subequations}
\label{eq:bte_php}
\begin{align}
\eta_{\lambda} + \gamma_\lambda^{-1} \partial_t \eta_{\lambda}
&= -\frac{\gamma_\lambda^{-1}}{k_\mathrm{B} T_0^2}
\vec{\Lambda}_\lambda \cdot \nabla T
-\gamma_\lambda^{-1}\sum_{\iota} \vec{\Pi}_{\lambda\iota}\cdot \nabla \theta_{\iota}, \\
\theta_{\iota}
&\approx -\varepsilon_\iota^{-1}\sum_{\lambda'} \vec{\Pi}_{\lambda'\iota}\cdot \nabla \eta_{\lambda'}.
\end{align}
\end{subequations}
Eliminating $\theta_\iota$ yields a closed equation for the hydrodynamic modes,
\begin{equation}
\label{eq:eta_hydro}
\big(1 + \gamma_\lambda^{-1} \partial_t \big) \eta_\lambda - \gamma_\lambda^{-1} (\overleftrightarrow{\mathcal N}\cdot \vec{\eta})_\lambda = -\frac{\gamma_\lambda^{-1}}{k_\mathrm{B} T_0^2} \vec{\Lambda}_\lambda \cdot \nabla T,
\end{equation}
with nonlocal operator
\begin{equation}
(\overleftrightarrow{\mathcal N}\cdot \vec{\eta})_\lambda = \sum_{\lambda'}\sum_{I,J} \mu_{\lambda\lambda' IJ}\,\partial^2_{IJ}\eta_{\lambda'}, \qquad \mu_{\lambda\lambda' IJ} = \sum_{\iota} \varepsilon_\iota^{-1} \Pi_{\lambda\iota,I}\Pi_{\lambda'\iota,J}.
\end{equation}
The operator $\overleftrightarrow{\mathcal N}$ represents nonlocal dissipation arising from the elimination of fast modes and generates viscous-like corrections to heat transport. The macroscopic heat flux is reconstructed as $\vec j = V^{-1}\sum_{\lambda=1}^{3} \vec{\Lambda}_\lambda\,\eta_\lambda$, where $\{\vec{\Lambda}_\lambda\}$ are linearly independent. Introducing dual vectors $\vec{\beta}_\lambda$ satisfying $\vec{\beta}_{\lambda'}\cdot\vec{\Lambda}_\lambda=\delta_{\lambda'\lambda}$ allows projection onto flux components and leads directly to the anisotropic hydrodynamic constitutive relation in the main text.

Crystal symmetry constrains $\mu_{\lambda\lambda' IJ}$. 
For example, in orthorhombic symmetry the operator reduces to
\begin{subequations}
\label{eq:L_orth}
\begin{align}
(\mathcal N\vec\eta)_1 &= (\mu_{1111}\partial_{xx}^2+\mu_{1122}\partial_{yy}^2+\mu_{1133}\partial_{zz}^2)\eta_1 + 2\mu_{1212}\partial_{xy}^2\eta_2 + 2\mu_{1313}\partial_{xz}^2\eta_3, \\
(\mathcal N\vec\eta)_2 &= (\mu_{2211}\partial_{xx}^2+\mu_{2222}\partial_{yy}^2+\mu_{2233}\partial_{zz}^2)\eta_2 + 2\mu_{2121}\partial_{xy}^2\eta_1 + 2\mu_{2323}\partial_{yz}^2\eta_3, \\
(\mathcal N\vec\eta)_3 &= (\mu_{3311}\partial_{xx}^2+\mu_{3322}\partial_{yy}^2+\mu_{3333}\partial_{zz}^2)\eta_3 + 2\mu_{3131}\partial_{xz}^2\eta_1 + 2\mu_{3232}\partial_{yz}^2\eta_2.
\end{align}
\end{subequations}
In the isotropic limit,
\begin{equation}
\mu_{\lambda\lambda'IJ} = \mu_{1122}\delta_{\lambda\lambda'}\delta_{IJ} +\frac12(\mu_{1111}-\mu_{1122}) (\delta_{\lambda I}\delta_{\lambda' J}+\delta_{\lambda J}\delta_{\lambda' I}),
\end{equation}
and the mode vectors reduce to Cartesian directions, recovering the standard Guyer--Krumhansl structure. With the constitutive PDE established, we now derive the equivalent spatiotemporal kernel.

\subsubsection{Spatiotemporal kernel representation}

The isotropic, curl-free G--K equation~\eqref{eq:gk_curl_free} admits a closed-form spatiotemporal kernel. Fourier--Laplace transforming ($\vec r\to\vec q$, $t\to s$) gives
\begin{equation}
\label{eq:GK_FLT}
(1 + \tau s + \tau\mu q^2)\,\hat{\tilde{j}}(\vec q,s) = -\kappa_0\,i\vec q\,\tilde T(\vec q,s),
\end{equation}
so the kernel in transform space is
\begin{equation}
\label{eq:GK_kernel_qs}
\tilde{\mathbb{Z}}_{\mathrm{PH}}(\vec q,s) = \frac{\kappa_0/\tau}{s + \tau^{-1} + \mu q^2}.
\end{equation}
Inverting the Laplace transform (simple pole at $s=-\tau^{-1}-\mu q^2$) yields
\begin{equation}
\label{eq:GK_kernel_qt}
\hat{\mathbb{Z}}_{\mathrm{PH}}(\vec q,t) = \frac{\kappa_0}{\tau}\,e^{-(1/\tau+\mu q^2)t}\,\Theta(t).
\end{equation}
The spatial dependence enters through the Gaussian factor $e^{-\mu q^2 t}$, whose three-dimensional inverse Fourier transform is $(4\pi\mu t)^{-3/2}\exp\!\bigl(-|\delta\vec r|^2/4\mu t\bigr)$ with $\delta\vec r=\vec R-\vec r$. Combining yields the PH kernel [Eq.~\eqref{eq:GK_kernel}]. The two asymptotic reductions stated in the main text then follow: spatial integration recovers the MCV kernel, while time integration via the identity $\int_0^\infty dt\,t^{-3/2}e^{-at-b/t}=({\pi}/{b})^{1/2}e^{-2\sqrt{ab}}$ yields the Yukawa-type ZD kernel. The full spatiotemporal integral recovers $\kappa_0$, confirming internal consistency with Fourier's law.

%%%%%%%%%%%%%%%%%%%%%%%%%%%%%%%%%%%%%%%%%%%%%%%%%%%%%%%%%%
\subsection{Attenuated Streaming Limit}
\label{app:AS}
%%%%%%%%%%%%%%%%%%%%%%%%%%%%%%%%%%%%%%%%%%%%%%%%%%%%%%%%%%

We derive here the attenuated-streaming (AS) limit of the Zwanzig-projected phBTE, corresponding to quasi-ballistic heat transport with finite propagation speed and temporal attenuation.

\subsubsection{Reduction}

The AS regime is obtained under a controlled reduction of the full projected dynamics in which each odd-parity heat-flux mode $\eta_\lambda$ is coupled predominantly to a single associated even-parity nonlocal mode $\theta_\lambda$. This \emph{paired-coupling} approximation neglects inter-mode mixing and retains only mode-diagonal coupling, $\Pi_{\lambda\iota} \rightarrow \Pi_{\lambda\lambda}\,\delta_{\lambda\iota}$. Physically, this approximation isolates the leading directional transport associated with each mode while suppressing higher-order collective interactions.

We further assume comparable relaxation scales for each paired mode, $\tau_\lambda \equiv \gamma_\lambda^{-1} \approx \varepsilon_\lambda^{-1}$, which is appropriate when the odd- and even-parity components of a given mode relax on similar time scales. Under these assumptions, the Zwanzig-projected equations reduce to
\begin{subequations}
\label{eq:AS_bte_pc}
\begin{align}
\big(1+\tau_\lambda \partial_t\big)\eta_\lambda(\vec r,t) &= -\frac{1}{k_\mathrm{B} T_0^2 \gamma_\lambda}\,\vec{\Lambda}_\lambda\cdot\nabla T(\vec r,t) -\vec{\ell}_\lambda\cdot\nabla \theta_\lambda(\vec r,t), \\
\big(1+\tau_\lambda \partial_t\big)\theta_\lambda(\vec r,t) &= -\vec{\ell}_\lambda\cdot\nabla \eta_\lambda(\vec r,t),
\end{align}
\end{subequations}
where the characteristic streaming vector is $\vec{\ell}_\lambda = \tau_\lambda\,\vec{\Pi}_{\lambda\lambda}$. Eliminating $\theta_\lambda$ from Eq.~\eqref{eq:AS_bte_pc} yields
\begin{equation}
\big(1+\tau_\lambda\partial_t\big)^2 \eta_\lambda - (\vec{\ell}_\lambda\cdot\nabla)^2 \eta_\lambda = -\frac{1}{k_\mathrm{B} T_0^2 \gamma_\lambda} \big(1+\tau_\lambda\partial_t\big)\,\vec{\Lambda}_\lambda\cdot\nabla T.
\end{equation}
Using  $\vec j_\lambda = V^{-1}\vec{\Lambda}_\lambda \eta_\lambda$  and $\overleftrightarrow{\kappa}_\lambda = \frac{\vec{\Lambda}_\lambda\otimes\vec{\Lambda}_\lambda}{V k_\mathrm{B} T_0^2 \gamma_\lambda}$, this becomes Eq. \eqref{eq:AS_constitutive_full}.
%\begin{equation}
%\label{eq:AS_constitutive_full}
%\big(1+\tau_\lambda\partial_t+\vec{\ell}_\lambda\cdot\nabla\big) \big(1+\tau_\lambda\partial_t-\vec{\ell}_\lambda\cdot\nabla\big) \vec j_\lambda = -\,\overleftrightarrow{\kappa}_\lambda\cdot\nabla\big(T+\tau_\lambda\partial_t T\big).
%\end{equation}
Because the left-hand side of Eq.~\eqref{eq:AS_constitutive_full} factorizes into two commuting first-order operators, it is natural to decompose $\vec j_\lambda = \vec j_{\lambda+} + \vec j_{\lambda-}$, with
\begin{subequations}
\label{eq:AS_split}
\begin{align}
\big(1+\tau_\lambda\partial_t+\vec{\ell}_\lambda\cdot\nabla\big)\vec j_{\lambda+}
&= -\frac{1}{2}\overleftrightarrow{\kappa}_\lambda\cdot\nabla T,
\\
\big(1+\tau_\lambda\partial_t-\vec{\ell}_\lambda\cdot\nabla\big)\vec j_{\lambda-}
&= -\frac{1}{2}\overleftrightarrow{\kappa}_\lambda\cdot\nabla T.
\end{align}
\end{subequations}
The operators $1+\tau_\lambda\partial_t \pm \vec{\ell}_\lambda\cdot\nabla$ therefore generate transport along straight-line characteristics with velocity $\vec{\ell}_\lambda/\tau_\lambda$ and exponential attenuation with time scale $\tau_\lambda$. Their causal Green's functions follow directly from the method of characteristics: $(1+\tau\partial_t+\vec{\ell}\cdot\nabla)G=\delta(t)\delta^{(3)}(\vec r)$ is solved by $G(\vec r,t)=\tau^{-1}e^{-t/\tau}\,\delta^{(3)}(\vec r-\vec{\ell}\,t/\tau)\,\Theta(t)$.

For a homogeneous medium, convolving these Green's functions with the source $-\frac{1}{2}\overleftrightarrow{\kappa}_\lambda\cdot\nabla T$ and summing over modes yields the AS kernel stated in the main text [Eq.~\eqref{eq:AS_kernel}]. This kernel describes propagation along $\pm \vec{\ell}_\lambda$ with finite speed and exponential attenuation. Spatial integration yields a local-transient kernel of MCV form, while time integration gives a steady nonlocal kernel with exponential spatial decay along the streaming direction.

The AS regime contains the phonon RTA as a limiting case. In the RTA, each phonon mode propagates ballistically with velocity $\vec v_\alpha$ between scattering events characterized by $\tau_\alpha$, yielding
\begin{equation}
\label{eq:rta_kernel_appendix}
\overleftrightarrow{\mathbb{Z}}_{\mathrm{RTA}}(\delta\vec r,t) = \sum_{\alpha} \frac{\overleftrightarrow{\kappa}_\alpha^{\mathrm{RTA}}}{\tau_\alpha} e^{-t/\tau_\alpha} \delta^{(3)}\!\big(\delta\vec r-\vec v_\alpha t\big)\Theta(t),
\end{equation}
where $\overleftrightarrow{\kappa}_\alpha^{\mathrm{RTA}} = \frac{1}{V}c_\alpha \tau_\alpha \,\vec v_\alpha\otimes\vec v_\alpha$. Thus, the RTA corresponds to the special case in which each phonon mode constitutes an independent attenuated-streaming channel. Spatial integration yields the local-transient RTA kernel,
\begin{equation}
\label{eq:rta_LT}
\overleftrightarrow{Z}_{\mathrm{RTA}}(t) = \sum_{\alpha} \frac{\overleftrightarrow{\kappa}^{\mathrm{RTA}}_\alpha}{\tau_\alpha} e^{-t/\tau_\alpha}\Theta(t),
\end{equation}
while time integration gives the steady spatial RTA kernel
\begin{equation}
\label{eq:rta_ZD}
\overleftrightarrow{Z}_{\mathrm{RTA}}(\delta\vec r) = \sum_{\alpha} \frac{\overleftrightarrow{\kappa}^{\mathrm{RTA}}_\alpha}{|\vec{v}_\alpha| \tau_\alpha} \delta^{(2)}(\delta\vec r_{\perp\alpha}) e^{-|s_\alpha|/(|\vec{v}_\alpha|\tau_\alpha)}.
\end{equation}
Substituting Eq.~\eqref{eq:rta_kernel_appendix} into the unified constitutive relation and carrying out the $\delta\vec r$ integration gives
\begin{equation}
\label{eq:RTA_constitutive_convected}
\vec j(\vec r,t) = -\sum_\alpha \frac{\overleftrightarrow{\kappa}^{\mathrm{RTA}}_\alpha}{\tau_\alpha} \int_0^\infty dt'\, e^{-t'/\tau_\alpha}\, \nabla T(\vec r-\vec v_\alpha t',\,t-t').
\end{equation}
Defining the mode-resolved flux 
\begin{equation}
\vec j_\alpha(\vec r,t) \equiv - \frac{\overleftrightarrow{\kappa}^{\mathrm{RTA}}_\alpha}{\tau_\alpha} \int_0^\infty dt'\, e^{-t'/\tau_\alpha}\, \nabla T(\vec r-\vec v_\alpha t',\,t-t'),
\end{equation}
and introducing the mean-free-path vector, $\vec \ell_\alpha \equiv \tau_\alpha \vec v_\alpha $,  one obtains
%\begin{equation}
%\left( 1+\tau_\alpha\partial_t+\tau_\alpha \vec v_\alpha\cdot\nabla \right)\vec j_\alpha(\vec r,t) = -\overleftrightarrow{\kappa}^{\mathrm{RTA}}_\alpha\cdot \nabla T(\vec r,t).
%\end{equation}
%Introducing the mean-free-path vector, this becomes
\begin{equation}
\label{eq:mode_constitutive_mfp}
\left( 1+\tau_\alpha\partial_t+\vec \ell_\alpha\cdot\nabla \right)\vec j_\alpha(\vec r,t) = -\overleftrightarrow{\kappa}^{\mathrm{RTA}}_\alpha\cdot \nabla T(\vec r,t).
\end{equation}

For homogeneous media, Fourier transformation gives
\begin{equation}
\label{eq:RTA_kernel_qw}
\overleftrightarrow{\widetilde{\mathbb Z}}_{\mathrm{RTA}}(\vec q,\omega) = \sum_{\alpha} \frac{\overleftrightarrow{\kappa}^{\mathrm{RTA}}_{\alpha}}{1+i\omega\tau_\alpha+i\,\vec q\cdot\vec \ell_\alpha},
\end{equation}
with directional propagator $(1+i\omega\tau_\alpha+i\,\vec q\cdot\vec \ell_\alpha)^{-1}$, which explicitly combines temporal relaxation and spatial streaming. Spatial integration of Eq.~\eqref{eq:rta_kernel_appendix} yields the local-transient kernel, while time integration gives the steady nonlocal kernel
\begin{equation}
\widetilde{\mathbb Z}^{\mathrm{ZD}}(\vec q) = \sum_\alpha \frac{\overleftrightarrow{\kappa}^{\mathrm{RTA}}_\alpha}{1+i\,\vec q\cdot\ell_\alpha},
\end{equation}
which is the $\omega=0$ limit of Eq.~\eqref{eq:RTA_kernel_qw}. In the joint limit \( |\omega\tau_\alpha|\ll1 \) and \( |\vec q\cdot\vec \ell_\alpha|\ll1, \) both reduce to Fourier transport.

\subsubsection{Relation to analytical Green's-function solutions of the phBTE}
\label{app:MH_relation}

The analytical Green's-function solution of the frequency-dependent phBTE derived by Hua and Minnich~\cite{hua2014analytical} provides a useful point of comparison with the RTA kernel formulation. Both approaches describe the same underlying semiclassical phonon transport physics, but they operate at different conceptual levels: the Hua--Minnich formulation gives a closed-form solution for the temperature field, whereas the present theory identifies the constitutive transport kernel itself as the fundamental object.

The same propagator structure appears in their solution before angular averaging. In their notation (adapted here to avoid conflict with the flux-mode variable $\eta_\lambda$), the deviational distribution function contains the denominator $1+i\Omega\tau_\omega+i\,\vec \xi\cdot\vec v\,\tau_\omega$, where $\Omega$ and $\vec\xi$ are temporal and spatial Fourier variables. This is directly analogous to the denominator in Eq.~\eqref{eq:RTA_kernel_qw}. For isotropic media, angular averaging transforms the directional propagator into $(\Lambda_\omega \xi)^{-1} \tan^{-1}\!\bigl( \Lambda_\omega \xi/(1+i\Omega\tau_\omega) \bigr)$, where $\Lambda_\omega=v_\omega\tau_\omega$ is the MFP. This scalar function appears explicitly in their temperature Green's function and represents the isotropically averaged counterpart of the directional RTA kernel.

The key distinction is therefore not the underlying physics, but the level of description. In the present formulation, $\overleftrightarrow{\mathbb Z}(\delta\vec r,t)$ directly characterizes the spatiotemporal constitutive response. In their formulation, one obtains the temperature Green's function after combining the phBTE with energy conservation and integrating over directions and frequencies. In this sense, the temperature Green's function may be viewed as a derived quantity obtained from the same underlying constitutive kernel through the continuity equation, $\rho c\,\frac{\partial T}{\partial t} + \nabla\cdot \vec j = Q$.

The kernel formulation therefore clarifies the relation between ballistic propagation, temporal memory, and spatial nonlocality at the constitutive level, while the Hua--Minnich solution provides the corresponding closed-form temperature response for homogeneous media after spectral and angular averaging.
%%%%%%%%%%%%%%%%%%%%%%%%%%%%%%%%%%%%%%%%%%%%%%%%%%%%%%%%%%
\subsection{Hybrid Collective--Streaming Reduction}
\label{app:hybrid}
%%%%%%%%%%%%%%%%%%%%%%%%%%%%%%%%%%%%%%%%%%%%%%%%%%%%%%%%%%

We derive here a reduced model that interpolates between the hydrodynamic and attenuated-streaming limits by retaining a subset of slow collective modes while treating the remaining modes in a paired streaming approximation. We partition the projected variables into a slow subset $S$ and a fast subset $F$. The slow subset contains modes with the longest relaxation times and is treated with full odd--even coupling. The remaining modes are treated using the paired-coupling approximation, $
\Pi_{\lambda\iota} \approx \Pi_{\lambda\lambda}\,\delta_{\lambda\iota}$, $\lambda\in F$.
Cross-couplings between $S$ and $F$ are neglected at leading order.

We first restrict the projected equations, Eqs.~\eqref{eq:bte_O_r} and \eqref{eq:bte_E_r}, to the retained slow subset. Let $\lambda,\lambda'\in S_{\mathrm O}$ denote the slow odd-parity modes and $\iota\in S_{\mathrm E}$ the corresponding slow even-parity modes. Neglecting couplings to the fast sector at leading order, the reduced equations are
\begin{subequations}
\label{eq:hybrid_slow_reduced}
\begin{align}
\partial_t \eta_\lambda + \gamma_\lambda \eta_\lambda &= -\frac{1}{k_\mathrm{B} T_0^2}\,\vec\Lambda_\lambda\cdot\nabla T -\sum_{\iota\in S_{\mathrm E}} \vec\Pi_{\lambda\iota}\cdot\nabla \theta_\iota, \label{eq:hybrid_slow_eta} \\
\partial_t \theta_\iota + \varepsilon_\iota \theta_\iota &= -\sum_{\lambda'\in S_{\mathrm O}} \vec\Pi_{\lambda'\iota}\cdot\nabla \eta_{\lambda'}. \label{eq:hybrid_slow_theta}
\end{align}
\end{subequations}

To obtain a closed equation for the slow odd modes, we eliminate the even-parity variables. In the collective sector we assume that the retained odd modes are the slow hydrodynamic-like variables, while the retained even modes remain slaved to them on the time scales of interest. To leading order, Eq.~\eqref{eq:hybrid_slow_theta} then gives
\begin{equation}
\label{eq:hybrid_theta_quasisteady}
\theta_\iota \approx -\varepsilon_\iota^{-1} \sum_{\lambda'\in S_{\mathrm O}} \vec\Pi_{\lambda'\iota}\cdot\nabla \eta_{\lambda'}.
\end{equation}
Substituting Eq.~\eqref{eq:hybrid_theta_quasisteady} into Eq.~\eqref{eq:hybrid_slow_eta} yields
\begin{equation}
\label{eq:hybrid_eta_closed}
\partial_t \eta_\lambda + \gamma_\lambda \eta_\lambda = -\frac{1}{k_\mathrm{B} T_0^2}\,\vec\Lambda_\lambda\cdot\nabla T + \sum_{\iota\in S_{\mathrm E}} \sum_{\lambda'\in S_{\mathrm O}} \vec\Pi_{\lambda\iota}\cdot \nabla\!\left[ \varepsilon_\iota^{-1} \vec\Pi_{\lambda'\iota}\cdot\nabla \eta_{\lambda'} \right].
\end{equation}

Expanding the spatial derivatives and keeping the leading second-order contribution gives
\begin{equation}
\label{eq:hybrid_eta_secondorder}
\partial_t \eta_\lambda + \gamma_\lambda \eta_\lambda = -\frac{1}{k_\mathrm{B} T_0^2}\,\vec\Lambda_\lambda\cdot\nabla T + \sum_{\lambda'\in S_{\mathrm O}} \sum_{I,J} \mu^{(S)}_{\lambda\lambda' IJ}\, \partial^2_{IJ}\eta_{\lambda'},
\end{equation}
with $\mu^{(S)}_{\lambda\lambda' IJ} = \sum_{\iota\in S_{\mathrm E}} \varepsilon_\iota^{-1} \Pi_{\lambda\iota,I}\Pi_{\lambda'\iota,J}$.

Equation~\eqref{eq:hybrid_eta_secondorder} is the analogue of the hydrodynamic reduction in Appendix~\ref{app:GK}, but restricted to the retained slow block. We now reconstruct the collective heat flux from the slow odd modes,
\begin{equation}
\label{eq:hybrid_flux_reconstruct}
\vec j^{(S)}(\vec r,t) = \frac{1}{V} \sum_{\lambda\in S_{\mathrm O}} \vec\Lambda_\lambda\,\eta_\lambda(\vec r,t).
\end{equation}
Introducing dual vectors $\vec\beta_\lambda$ satisfying $
\vec\beta_{\lambda'}\cdot\vec\Lambda_\lambda=\delta_{\lambda'\lambda}$,
and multiplying Eq.~\eqref{eq:hybrid_eta_secondorder} by $V^{-1}\vec\Lambda_\lambda$, one obtains a closed constitutive equation for $\vec j^{(S)}$ of the form
\begin{equation}
\label{eq:hybrid_collective_appendix}
\bigl(1+\tau^{(S)}\partial_t\bigr)\vec j^{(S)} - \tau^{(S)}\mathcal{D}^{(S)}\vec j^{(S)} = -\,\overleftrightarrow{\kappa}^{(S)}\cdot\nabla T.
\end{equation}
Here $\tau^{(S)}$ and $\overleftrightarrow{\kappa}^{(S)}$ denote the effective relaxation time and conductivity tensor associated with the retained slow block, and the second-order spatial operator is
\begin{equation}
\label{eq:hybrid_D_operator}
\mathcal{D}^{(S)} \vec j^{(S)} = \sum_{\lambda,\lambda'\in S_{\mathrm O}} \sum_{I,J} \mu^{(S)}_{\lambda\lambda' IJ}\, \partial^2_{IJ}\vec j_{\lambda'}^{(S)}.
\end{equation}

This recovers the hydrodynamic reduction when $S_{\mathrm O}$ consists of the three hydrodynamic modes. More generally, it describes a collective subspace of arbitrary finite dimension, embedded within a broader transport spectrum that is treated separately by the streaming sector.

We next consider the complementary fast sector, consisting of the modes not retained in the collective subset. For these modes, we adopt the paired-coupling approximation introduced in Appendix~\ref{app:AS}: each odd-parity mode $\eta_\lambda$ is coupled predominantly to a single even-parity partner $\theta_\lambda$, while intermode mixing within the fast sector is neglected. The reduced equations are therefore
\begin{subequations}
\label{eq:hybrid_fast_reduced}
\begin{align}
\partial_t \eta_\lambda + \gamma_\lambda \eta_\lambda &= -\frac{1}{k_\mathrm{B} T_0^2}\,\vec\Lambda_\lambda\cdot\nabla T -\vec\Pi_{\lambda\lambda}\cdot\nabla \theta_\lambda, \label{eq:hybrid_fast_eta} \\
\partial_t \theta_\lambda + \varepsilon_\lambda \theta_\lambda &= -\vec\Pi_{\lambda\lambda}\cdot\nabla \eta_\lambda, \label{eq:hybrid_fast_theta}
\end{align}
\end{subequations}
where now $\lambda\in F$ labels the fast paired modes.

As in the attenuated-streaming limit, we assume that the odd- and even-parity members of a given fast pair relax on comparable time scales,
\begin{equation}
\label{eq:hybrid_tau_fast}
\tau_\lambda \equiv \gamma_\lambda^{-1}\approx \varepsilon_\lambda^{-1}, \qquad \vec\ell_\lambda \equiv \tau_\lambda\,\vec\Pi_{\lambda\lambda}.
\end{equation}
Multiplying Eqs.~\eqref{eq:hybrid_fast_eta} and \eqref{eq:hybrid_fast_theta} by $\tau_\lambda$ then gives
\begin{subequations}
\label{eq:hybrid_fast_scaled}
\begin{align}
\bigl(1+\tau_\lambda\partial_t\bigr)\eta_\lambda&=-\frac{1}{k_\mathrm{B} T_0^2\gamma_\lambda}\,\vec\Lambda_\lambda\cdot\nabla T-\vec\ell_\lambda\cdot\nabla\theta_\lambda,
\label{eq:hybrid_fast_scaled_eta}
\\
\bigl(1+\tau_\lambda\partial_t\bigr)\theta_\lambda&=-\vec\ell_\lambda\cdot\nabla\eta_\lambda.
\label{eq:hybrid_fast_scaled_theta}
\end{align}
\end{subequations}
To eliminate the even-parity variable, we apply the operator
$\bigl(1+\tau_\lambda\partial_t\bigr)$ to Eq.~\eqref{eq:hybrid_fast_scaled_eta} and substitute Eq.~\eqref{eq:hybrid_fast_scaled_theta}. This yields
\begin{equation}
\label{eq:hybrid_fast_eta_closed}
\bigl(1+\tau_\lambda\partial_t\bigr)^2\eta_\lambda-(\vec\ell_\lambda\cdot\nabla)^2\eta_\lambda=-\frac{1}{k_\mathrm{B} T_0^2\gamma_\lambda}\bigl(1+\tau_\lambda\partial_t\bigr)\vec\Lambda_\lambda\cdot\nabla T.
\end{equation}
Because the differential operators commute, the left-hand side factorizes as
\begin{equation}
\label{eq:hybrid_fast_factorized_eta}
\bigl(1+\tau_\lambda\partial_t+\vec\ell_\lambda\cdot\nabla\bigr)\bigl(1+\tau_\lambda\partial_t-\vec\ell_\lambda\cdot\nabla\bigr)\eta_\lambda=-\frac{1}{k_\mathrm{B} T_0^2\gamma_\lambda}\bigl(1+\tau_\lambda\partial_t\bigr)\vec\Lambda_\lambda\cdot\nabla T.
\end{equation}

We reconstruct the modal heat flux as $\vec j_\lambda = V^{-1}\,\vec\Lambda_\lambda\,\eta_\lambda$, with the corresponding conductivity tensor $\overleftrightarrow{\kappa}_\lambda = \vec\Lambda_\lambda \otimes \vec\Lambda_\lambda / (V k_\mathrm{B} T_0^2 \gamma_\lambda)$. This yields the streaming-sector constitutive relation in the form of Eq.~\eqref{eq:AS_constitutive_full}.

As shown in Appendix~\ref{app:AS}, the factorized structure admits a directional decomposition, Eqs.~\eqref{eq:AS_split}, in which $\vec j_\lambda = \vec j_{\lambda+} + \vec j_{\lambda-}$. These two first-order equations describe transport along the characteristics
$\pm \vec\ell_\lambda/\tau_\lambda$, with exponential attenuation on the time scale $\tau_\lambda$. The total fast-sector flux is therefore $
\vec j^{(F)}(\vec r,t)=\sum_{\lambda\in F}\bigl(\vec j_{\lambda+}+\vec j_{\lambda-}\bigr)$, which is the form used in the main text.

Both sectors admit kernel representations. The collective sector yields a Gaussian-type hydrodynamic kernel,
\begin{equation}
\overleftrightarrow{\mathbb Z}_{\mathrm{PH}}^{(S)} = \frac{\overleftrightarrow{\kappa}^{(S)}}{\tau^{(S)}} \frac{e^{-t/\tau^{(S)}}}{(4\pi \mu^{(S)} t)^{3/2}} \exp\!\left(-\frac{|\delta\vec r|^2}{4\mu^{(S)} t}\right)\Theta(t),
\end{equation}
while the streaming sector yields
\begin{equation}
\overleftrightarrow{\mathbb Z}_{\mathrm{AS}}^{(F)} = \sum_{\lambda\in F} \frac{\overleftrightarrow{\kappa}_\lambda}{2\tau_\lambda} e^{-t/\tau_\lambda} \Big[ \delta^{(3)}(\delta\vec r-\vec v_\lambda t) + \delta^{(3)}(\delta\vec r+\vec v_\lambda t) \Big]\Theta(t).
\end{equation}

The hybrid kernel is therefore
\begin{equation}
\overleftrightarrow{\mathbb Z}_{\mathrm{HCS}} = \overleftrightarrow{\mathbb Z}_{\mathrm{PH}}^{(S)} + \overleftrightarrow{\mathbb Z}_{\mathrm{AS}}^{(F)},
\end{equation}
which provides a controlled interpolation between collective hydrodynamic transport and quasi-ballistic streaming in systems with broad spectral distributions.

\section{Full Spatiotemporal Kernels}
\label{app:full_kernel}

\subsection{Unified Constitutive Relation }
\label{app:unified}

The iterative expansion [Eq.~\eqref{eq:J_expansion_r}] can be resummed into the unified spatiotemporal constitutive relation [Eq.~\eqref{eq:j_convolution}]. Using the Taylor expansion [Eq. \eqref{eq:vector_taylor}], the single-time kernels $\overleftrightarrow{Z}^{(n)}$ appearing in Eq.~\eqref{eq:J_expansion_r} are related to the full kernel $\overleftrightarrow{\mathbb Z}$ through spatial moments,
\begin{equation}
\label{eq:Y_moment}
\overleftrightarrow{Z}^{(n_1+n_2+n_3)}(\vec{r}, t; n_1, n_2, n_3) = \int_\mathbb{V} d^3\vec{R}\, \overleftrightarrow{\mathbb{Z}}(\vec{r}, \vec{R}, t) \, {(\delta x)}^{n_1} {(\delta y)}^{n_2} {(\delta z)}^{n_3}.
\end{equation}
This moment relation provides the bridge between the iterative gradient expansion and the unified constitutive kernel. For an inhomogeneous phonon gas, the constitutive structure emerges directly from the Zwanzig-projected phBTE: eliminating the fast microscopic degrees of freedom yields a causal, non-Markovian relation whose temporal memory and spatial nonlocality are encoded in well-defined convolution kernels determined by phonon dispersions, scattering rates, and mode couplings.

\subsection{Time-Domain vs Frequency-Domain Representations}
\label{sec:time_frequency}

The time-domain constitutive relation may be expressed equivalently in the frequency domain as
\begin{equation}
\label{eq:Fourier_law_in_omega}
\tilde{\vec j}(\vec r,\omega) = -\int_{\mathbb V} d^3\vec R\; \overleftrightarrow{\tilde{\mathbb{Z}}}(\vec r,\vec R,\omega) \cdot \nabla \tilde T(\vec R,\omega),
\end{equation}
with
\begin{subequations}
\begin{align}
\tilde{\vec j}(\vec r,\omega) &= \int_0^\infty dt\, \vec j(\vec r,t)\,e^{i\omega t},
\\
\nabla \tilde T(\vec R,\omega) &= \int_0^\infty dt\, \nabla T(\vec R,t)\,e^{i\omega t},
\\
\overleftrightarrow{\tilde{\mathbb{Z}}}(\vec r,\vec R,\omega) &= \int_0^\infty dt\, \overleftrightarrow{\mathbb{Z}}(\vec r,\vec R,t)\,e^{i\omega t}.
\end{align}
\end{subequations}
Equations~\eqref{eq:j_convolution} and \eqref{eq:Fourier_law_in_omega} are mathematically equivalent.

While frequency-domain response functions are natural for externally driven electronic or optical transport, heat conduction is driven by internally generated thermodynamic gradients. The evolution of the temperature field therefore reflects intrinsic relaxation dynamics, making the time-domain formulation especially transparent for dissipation processes and transient thermal measurements.

\subsection{Controlled Asymptotic Limits}
\label{app:limits}
The lowest-order term, $n_1=n_2=n_3=0$, yields the local--transient limit,
\begin{subequations}
\label{eq:LT_reduction}
\begin{eqnarray}
\overleftrightarrow{Z}^{(0)}(\vec{r},t';0,0,0)= \overleftrightarrow{Z}_{\mathrm{LT}}(\vec r,t) = \int_{\mathbb V} d^3\vec R\; \overleftrightarrow{\mathbb{Z}}(\vec r,\vec R,t),
\\
\partial_t T(\vec r,t) - \int_{0}^{\infty} dt'\; \nabla \cdot \Big[ \frac{\overleftrightarrow{Z}_{\mathrm{LT}}(\vec r,t')}{\rho(\vec r)c(\vec r)} \cdot \nabla T(\vec r,t-t') \Big] =0.
\end{eqnarray}
\end{subequations}
This regime retains temporal memory while neglecting spatial nonlocality.

When temperature variations evolve slowly relative to intrinsic relaxation times, temporal memory becomes negligible and the time-integrated spatial kernel fully characterizes nonlocal diffusion,
\begin{subequations}
\label{eq:ZD_reduction}
\begin{eqnarray}
\overleftrightarrow{Z}_{\mathrm{ZD}}(\vec r,\vec R) = \int_{0}^{\infty} dt\; \overleftrightarrow{\mathbb{Z}}(\vec r,\vec R,t),
\\
\partial_t T(\vec r,t) - \int_{\mathbb V} d^3\vec R\; \nabla \cdot \Big[ \frac{\overleftrightarrow{Z}_{\mathrm{ZD}}(\vec r,\vec R)}{\rho(\vec r)c(\vec r)} \cdot \nabla T(\vec R,t) \Big] =0.
\end{eqnarray}
\end{subequations}
This is the Zwanzig diffusion limit.

Spatial kernel descriptions of heat conduction have appeared previously in phenomenological studies, but typically as empirical fitting forms restricted to homogeneous media. By contrast, the present spatial diffusion kernel is derived directly from the Zwanzig projection-operator formalism and therefore admits a clear microscopic interpretation. For media with smoothly varying properties, the same framework naturally accommodates nanostructures, confined geometries, and graded materials through explicit position dependence of the kernel. Sharp material interfaces, where phonon modes change abruptly, require additional treatment via the Dyson-type construction developed in Sec.~\ref{sec:generalized_kapitza}.

Higher-order spatial nonlocality may equivalently be characterized by generalized conductivity tensors defined as the spatial moments of the kernel,
\begin{equation}
\label{eq:Z_to_general_kappa}
\overleftrightarrow{\kappa}^{(n_1+n_2+n_3)}(\vec{r};n_1,n_2,n_3) = \int_{\mathbb{V}} d^3\vec{R} \int_{0}^{\infty} dt\; \overleftrightarrow{\mathbb{Z}}(\vec{r},\vec{R},t) (\delta x)^{n_1}(\delta y)^{n_2}(\delta z)^{n_3},
\end{equation}
which quantify progressively higher-order departures from locality.

When both the spatial range and memory time are small relative to macroscopic scales, the formulation reduces to classical Fourier diffusion,
\begin{subequations}
\label{eq:LT-ZD_reduction}
\begin{eqnarray}
\overleftrightarrow{\kappa}(\vec r) &=& \int_{\mathbb V} d^3\vec R \int_{0}^{\infty} dt\; \overleftrightarrow{\mathbb{Z}}(\vec r,\vec R,t),\quad \overleftrightarrow{D}(\vec r)= \frac{\overleftrightarrow{\kappa}(\vec r)} {\rho(\vec r)c(\vec r)} 
\\
\partial_t T(\vec r,t) &-& \nabla \cdot \Big[ \overleftrightarrow{D}(\vec r) \cdot \nabla T(\vec r,t) \Big] =0.
\end{eqnarray}
\end{subequations}
Classical diffusion is therefore recovered as a double asymptotic limit of the full spatiotemporal kernel description.

In summary, the kernel formulation provides a unified description of transient, nonlocal, and classical heat conduction under the fundamental assumption that a temperature field exists. Fully ballistic transport without local equilibration lies outside this thermodynamic closure and must instead be treated kinetically.

%%%%%%%%%%%%%%%%%%%%%%%%%%%%%%%%%%%%%%%%%%%%%%%%
\section{Derivation of the Generalized Green--Kubo formula}
\label{app:Robertson_Z}

In this Appendix, we derive the spatiotemporal heat-conduction kernel $\overleftrightarrow{\mathbb Z}(\vec r,\vec R,t)$ as a space-projected heat-flux correlation function within nonequilibrium statistical mechanics. The derivation is based on the Zwanzig projection-operator formalism for irreversible processes \cite{zwanzig1960ensemble,zwanzig1961memory,zwanzig1964elementary} together with Robertson’s nonlinear extension for spatially inhomogeneous systems \cite{robertson1966,robertson1967}. 

Zubarev’s nonequilibrium statistical operator (NSO) method \cite{zubarev1992nonequilibrium} yields equivalent constitutive relations for transient and nonlocal heat conduction when formulated with the same relevant variables and conservation laws. Both Robertson’s projection framework and Zubarev’s NSO construction build nonequilibrium states from a local-equilibrium reference supplemented by causal corrections, implemented through projection operators in Robertson’s case and retarded source terms in Zubarev’s formulation. When applied consistently, both approaches produce identical flux-force relations.

\subsection{Projected Dynamics in the Local Thermal Equilibrium Limit}

Let $\hat f(t)$ denote the phase-space probability density operator evolving under the Liouville equation, $\partial_t \hat f = -i\hat{\mathcal L}\hat f$, where $\hat{\mathcal L}$ is the Liouvillian operator. A projection operator $\hat{\mathcal P}$ separates \emph{relevant} and \emph{irrelevant} components of the dynamics, defining the relevant probability density operator $\hat f_R(\Gamma,t)=\hat{\mathcal P}\,\hat f(\Gamma,t)$. Projecting the Liouville equation onto the $\hat{\mathcal{P}}$ subspace and imposing the retarded condition $\delta f(t\to-\infty)=0$ yields a closed equation of motion for $f_R$ \cite{zwanzig1960ensemble},
\begin{equation}
\label{eq:f_R}
\left(\partial_t+i\,\hat{\mathcal{P}}\hat{\mathcal{L}}\hat{\mathcal{P}}\right) f_R(t) = -\int_{-\infty}^{t} ds\,\hat{\mathcal{K}}(t-s)\,f_R(s),
\end{equation}
with the memory kernel operator, 
\begin{equation}
\label{eq:fR_kernel}
\hat{\mathcal{K}}(t)= \hat{\mathcal{P}}\hat{\mathcal{L}}\hat{\mathcal{Q}}\, e^{-it\hat{\mathcal{Q}}\hat{\mathcal{L}}\hat{\mathcal{Q}}}\, \hat{\mathcal{Q}}\hat{\mathcal{L}}\hat{\mathcal{P}},
\end{equation}
encoding the statistical influence of the irrelevant variables.

For heat transport, we select the local energy-density operator $\hat h(\vec r)$ as the relevant observable. Microscopic energy conservation, $\partial_t\hat h(\vec r,t) + \nabla\!\cdot\!\vec{\hat j}(\vec r,t) = 0$, implies that the microscopic heat flux emerges directly from projected dynamics of $\hat{h}(\vec r)$,
\begin{equation}
\label{eq:local_energy_conservation}
\nabla\!\cdot\!\vec{\hat j}(\vec r) = i\hat{\mathcal L}\hat h(\vec r).
\end{equation}

Following Robertson’s nonlinear extension to spatially inhomogeneous systems, we introduce a time-dependent projection operator $\hat{\mathcal P}(t)$ associated with a relevant distribution $\hat f_R(t)$. The key physical approximation is the local thermal equilibrium (LTE) assumption, under which the nonequilibrium ensemble is represented by a generalized canonical form,
\begin{equation}
\hat f_R(t) = \exp\!\left[-\Phi(t) - \int d\vec r\, \frac{\hat h(\vec r)}{k_\mathrm{B} T(\vec r,t)} \right],
\end{equation}
where the temperature field $T(\vec r,t)$ is determined self-consistently. In the homogeneous limit $T(\vec r,t)\to T_0$, this reduces to the equilibrium distribution $\hat f_{eq}=e^{-\hat{\mathcal H}/(k_\mathrm{B} T_0)}/Z_{eq}$, with the Hamiltonian $\hat{\mathcal H}=\int d\vec r\,\hat h(\vec r)$.

Combining the time-dependent projection operator with the generalized canonical distribution yields an integro-differential equation for the expectation value of the local energy density,\( e(\vec r,t)=\langle \hat h(\vec r)\rangle_t \),
\begin{equation}
\label{eq:robertson_eom}
\partial_t e(\vec r,t) = \int_{-\infty}^{t} dt'\! \int_V d\vec R\, K(\vec r,t;\vec R,t') \,[k_\mathrm{B} T(\vec R,t')]^{-1},
\end{equation}
where $\langle\cdots\rangle_t$ denotes averaging with respect to $\hat{f}_R(t)$ and the kernel is given by
\begin{equation}
\label{eq:robertson_kernel}
K(\vec r,t;\vec R,t') = \Big\langle [i\hat{\mathcal L}\hat h(\vec r)]\, \hat{\mathcal T}(t,t')\, [1-\hat{\mathcal P}(t')]\, \overline{[i\hat{\mathcal L}\hat h(\vec R)]} \Big\rangle_t .
\end{equation}
Here $\hat{\mathcal T}(t,t')$ is the orthogonal propagator defined by
\begin{equation}
\label{eq:robertson_propagator}
\partial_{t'}\hat{\mathcal T}(t,t') = \hat{\mathcal T}(t,t')\, [1-\hat{\mathcal P}(t')]\, i\hat{\mathcal L},
\end{equation}
and
\begin{equation}
\overline{\hat A} = \int_{0}^{1} d\zeta\, \hat{f}_R^\zeta(t)\,\hat A\,\hat{f}_R^{1-\zeta}(t) - \langle \hat A\rangle_t .
\end{equation}

Equations~\eqref{eq:robertson_eom}--\eqref{eq:robertson_kernel} show that transport in spatially inhomogeneous systems is intrinsically nonlocal in space and retarded in time, with both effects encoded in $K(\vec r,t;\vec R,t')$.
Using Eq.~\eqref{eq:local_energy_conservation}, Eq.~\eqref{eq:robertson_eom} can be rewritten in constitutive form,
\begin{equation}
\label{eq:robertson_jkernel}
\vec j(\vec r,t) = -\int_V d\vec R \int_{-\infty}^{t} dt'\, \overleftrightarrow{\mathbb Z}(\vec r,t;\vec R,t') \cdot \nabla T(\vec R,t'),
\end{equation}
where $\vec j(\vec r,t)=\langle\vec{\hat j}(\vec r)\rangle_t$ and
\begin{equation}
\label{eq:robertson_Zkernel}
\overleftrightarrow{\mathbb Z}(\vec r,t;\vec R,t') = \frac{1}{k_\mathrm{B} T_0^2} \Big\langle \vec{\hat j}(\vec r) \otimes \hat{\mathcal T}(t,t')\, [1-\hat{\mathcal P}(t')]\, \overline{\vec{\hat j}(\vec R)} \Big\rangle_t .
\end{equation}

%%%%%%%%%%%%%%%%%%%%%%%%%%%%%%%%%%%%%%%%%%%%%%%%%
\subsection{Linear Response}
\label{app:linear_response_hardy_J}
Equation~\eqref{eq:robertson_jkernel} is formally exact within the LTE assumption. 
To connect this result with conventional Green--Kubo expressions, we linearize about global equilibrium. In this limit,
\begin{equation}
\hat{f}_{R}(t) \simeq \hat f_{eq} \left[ 1 - \frac{1}{k_\mathrm{B} T_0^2} \int d\vec R\, \delta T(\vec R,t) \left( \hat h(\vec R) - \langle \hat h(\vec R)\rangle_{eq} \right) \right],
\end{equation}
and the orthogonal propagator reduces to equilibrium time evolution. Under this approximation,
\begin{equation}
\label{eq:robertson_short_time}
\Big\langle \vec{\hat j}(\vec r) \otimes \hat{\mathcal T}(t,t')\, [1-\hat{\mathcal P}(t')]\, \overline{\vec{\hat j}}(\vec R) \Big\rangle_t \simeq \langle \vec j(\vec r,t) \otimes \vec j(\vec R,t') \rangle_{eq},
\end{equation}
where $\langle\cdots\rangle_{eq}$ denotes averaging with respect to $\hat f_{eq}$. The spatiotemporal kernel then reduces to the generalized Green--Kubo form [Eq.~\eqref{eq:gGK_full}], providing a direct microscopic interpretation of the kernel as a space-resolved heat-flux time-correlation function.

Structurally, Eqs.~\eqref{eq:robertson_jkernel} and \eqref{eq:gGK_full} are identical to the spatiotemporal constitutive relations derived in the Zwanzig-projected phonon Boltzmann formulation developed in the main text. The distinction lies in the level of microscopic specification: for phonon gases, the kernel can be expressed explicitly in terms of phonon mode properties and scattering operators, whereas Robertson’s formulation defines the kernel formally for arbitrary interacting many-body systems. This establishes Robertson’s theory as a universal statistical-mechanical foundation for material-specific spatiotemporal kernels and provides the general microscopic basis for the kernel formalism employed throughout this work.

\subsection{Bond-Centered Heat-Flux Operator}
\label{app:hardy_bond_flux}

We now construct the spatially resolved heat-flux operator entering the generalized Green--Kubo relation [Eq.~\eqref{eq:gGK_full}] and use it to derive the spatial conductivity kernel for disordered harmonic solids. The derivation proceeds in two steps: we first obtain a bond-centered representation of the local heat flux following Hardy~\cite{hardy1963energy}, and then evaluate the corresponding space-projected heat-flux correlation function.

The microscopic conductive heat-flux density can be written as~\cite{hardy1963energy},
\begin{equation}
\label{eq:Hardy_eq2-12_app}
\vec{j}(\vec r) = \frac{1}{2} \sum_{ij} \vec{R}_{ij} \left( 1 + \frac{1}{2!}\vec{R}_{ij}\!\cdot\!\nabla + \frac{1}{3!}(\vec{R}_{ij}\!\cdot\!\nabla)^2 + \cdots \right) \Delta(\vec r-\vec R_i)\,\Xi_{ij} + \mathrm{H.c.},
\end{equation}
where $\vec R_{ij}=\vec R_i-\vec R_j$, and $\Delta(\vec r-\vec R_i)$ is a localization function satisfying $\int d\vec r\, \Delta(\vec r-\vec R_i)=1$. The operator $\Xi_{ij} = (i\hbar)^{-1}[ p_i^2/2m_i,\, V_j ]$ represents pairwise energy transfer. The differential series can be resummed using the identity $\int_0^1 d\xi\, e^{\xi A} = \sum_{n=0}^{\infty} A^n/(n+1)!$, giving $\vec j(\vec r) = \tfrac{1}{2} \sum_{ij} \vec R_{ij}\,\Xi_{ij} \int_0^1 d\xi\, e^{\xi\vec R_{ij}\cdot\nabla} \Delta(\vec r-\vec R_i) + \mathrm{H.c.}$ Using the translation identity $e^{\vec a\cdot\nabla} f(\vec r)=f(\vec r+\vec a)$, we obtain $\Delta(\vec r-\vec R_i+\xi\vec R_{ij}) = \Delta\!\big(\vec r-[(1-\xi)\vec R_i+\xi\vec R_j]\big)$.

Defining the Hermitian bond operator and projection function, $
\Psi_{ij}=\frac{1}{2}(\Xi_{ij}+\Xi_{ij}^\dagger)$ and $w(\vec r;\vec R_i,\vec R_j) = \int_0^1 d\xi\, \Delta\!\big(\vec r-[(1-\xi)\vec R_i+\xi\vec R_j]\big)$, the local heat flux takes the bond-centered form [Eq.~\eqref{eq:j_bond}], in which each term contributes a flux directed along the bond $\vec R_{ij}$, distributed over the interatomic segment by the weight $w$. Spatial integration recovers the bulk current $\vec j_{\mathrm{bulk}} = V^{-1}\sum_{ij} \Psi_{ij}\cdot(\vec R_j-\vec R_i)$, independent of the localization function.

%%%%%%%%%%%%%%%%%%%%%%%%%%%%%%%%%%%%%%%%%%%%%%%%%%%%%%%%%%
\subsection{Spatial kernel for disordered harmonic lattices}
\label{app:AF_kernel}
%%%%%%%%%%%%%%%%%%%%%%%%%%%%%%%%%%%%%%%%%%%%%%%%%%%%%%%%%%

We now use the bond-centered flux operator [Eq.~\eqref{eq:j_bond}] to evaluate the space-projected heat-flux correlation function in the harmonic limit in a disordered solid. This yields a spatial conductivity kernel that reduces, in the Zwanzig diffusion limit, to the Allen--Feldman description of thermal transport in disordered solids~\cite{Allen_Feldman_PhysRevB.48.12581}. For a harmonic lattice, the bond operator becomes
\begin{equation}
\Psi_{ij}(t) = \frac{1}{2} \left[ \dot{\vec u}_i\cdot\mathbf{\Phi}_{ij}\cdot \vec u_j - \vec u_i\cdot\mathbf{\Phi}_{ij}\cdot \dot{\vec u}_j \right],
\end{equation}
leading to $\vec j(\vec r,t) = -\sum_{ij} \vec R_{ij}\, w(\vec r;\vec R_i,\vec R_j)\, \Psi_{ij}(t)$. Expanding in normal modes, $\vec u_i = m_i^{-1/2}\sum_s \vec e_i^{(s)} Q_s$, gives the bilinear form $\vec j(\vec r,t) = \sum_{ss'} \vec{\mathcal{J}}_{ss'}(\vec r)\, (Q_s\dot Q_{s'}-\dot Q_s Q_{s'})$. In second quantization and retaining number-conserving terms, $\vec j(\vec r,t) = \sum_{ss'} \vec{\mathcal{S}}_{ss'}(\vec r)\, a_s^\dagger a_{s'}\, e^{i(\omega_s-\omega_{s'})t}$.

The spatial conductivity kernel is defined as
\begin{equation}
\label{eq:ZZD_GK_def}
\overleftrightarrow{Z}_{\mathrm{ZD}}(\vec r,\vec R) = \frac{1}{k_\mathrm{B} T^2} \int_0^\infty dt\, \langle \vec j(\vec r,t)\otimes \vec j(\vec R,0)\rangle.
\end{equation}

Evaluating the harmonic correlation function yields
\begin{equation}
\label{eq:ZZD_AF_spatial}
\overleftrightarrow{Z}_{\mathrm{ZD}}(\vec r,\vec R) = \pi \sum_s \frac{C_s}{\omega_s^2} \sum_{s'\neq s} \vec{\mathcal{S}}_{ss'}(\vec r)\otimes \vec{\mathcal{S}}_{s's}(\vec R)\, \delta(\omega_s-\omega_{s'}).
\end{equation}
This expression shows that thermal transport is governed by resonant coupling between nearly degenerate vibrational modes, with the $\delta$-function enforcing the frequency-matching condition central to Allen--Feldman theory.

The bulk conductivity follows from spatial integration, $\overleftrightarrow{\kappa} = V^{-1}\int d\vec r\,d\vec R\, \overleftrightarrow{Z}_{\mathrm{ZD}}(\vec r,\vec R)$, giving $\overleftrightarrow{\kappa}_{\mathrm{AF}} = \frac{1}{V}\sum_s C_s \overleftrightarrow{D}_s$ and  $\overleftrightarrow{D}_s = \pi \sum_{s'\neq s} \frac{\vec S_{ss'}\otimes \vec S_{s's}}{\omega_s^2} \delta(\omega_s-\omega_{s'})$. Thus, the Allen--Feldman result emerges as the spatially integrated limit of the nonlocal kernel. The present formulation retains the full spatial structure of mode coupling, providing a real-space generalization of harmonic thermal transport. Finally, while the spatial decomposition depends on the projection function $w(\vec r;\vec R_i,\vec R_j)$, the integrated conductivity is gauge invariant and coincides with the conventional Allen--Feldman result.

%\newpage
 
\bibliography{unified_theory}% Produces the bibliography via BibTeX.

\end{document}